\documentclass[aps,prb,twocolumn,superscriptaddress,showpacs,floatfix,longbibliography]{revtex4-2}
\usepackage{fullpage}
\usepackage{mathrsfs} \usepackage{amssymb} \usepackage{graphicx,color} \usepackage{bbm} \usepackage{multirow}
\usepackage{bm} \usepackage{mathrsfs} \usepackage{commath} \usepackage{stmaryrd} \usepackage{color}
\usepackage{subfigure}
\usepackage{comment}
\usepackage{inputenc}
\usepackage[normalem]{ulem}
\usepackage{amsmath}
\usepackage{amssymb}
\usepackage{bm}
\usepackage{graphicx}
\usepackage{float}

\usepackage{hyperref}
\hypersetup{
  colorlinks   = true,    % Colours links instead of ugly boxes
  urlcolor     = blue,    % Colour for external hyperlinks
  linkcolor    = blue,    % Colour of internal links
  citecolor    = blue     % Colour of citations
}

\newcommand{\beq}{\begin{equation}}
\newcommand{\eeq}{\end{equation}}

\begin{document}

\title{ Antiferromagnetic order enhanced by local dissipation
}

\author{Oscar Bouverot-Dupuis}
\affiliation{Universit\'{e} Paris Saclay, CNRS, LPTMS, 91405, Orsay, France}
\affiliation{IPhT, CNRS, CEA, Universit\'{e} Paris Saclay, 91191 Gif-sur-Yvette, France}

\author{Saptarshi Majumdar}
\affiliation{Universit\'{e} Paris Saclay, CNRS, LPTMS, 91405, Orsay, France}
\affiliation{IPhT, CNRS, CEA, Universit\'{e} Paris Saclay, 91191 Gif-sur-Yvette, France}

\author{Alberto Rosso}
\affiliation{Universit\'{e} Paris Saclay, CNRS, LPTMS, 91405, Orsay, France}

\author{Laura Foini}
\affiliation{IPhT, CNRS, CEA, Universit\'{e} Paris Saclay, 91191 Gif-sur-Yvette, France}

\begin{abstract}
We study an XXZ spin chain at zero magnetization coupled to a collection of local harmonic baths at zero temperature. We map this system on a (1+1)D effective field theory using bosonization, where the effect of the bath is taken care of in an exact manner. We provide analytical and numerical evidence of the existence of two phases at zero temperature: a Luttinger liquid (LL) and an antiferromagnetic phase (AFM), separated by a phase transition akin to the Berezinsky--Kosterlitz--Thouless (BKT) type. While the bath is responsible for the LL-AFM transition for subohmic baths, the LL-AFM transition for superohmic baths is due to the interactions within the spin chain.
\end{abstract}

\date{\today}

\maketitle

\section{Introduction}
The problem of a quantum-mechanical system interacting with an environment is ubiquitous in physics. These interactions can originate from a surrounding bath, an external driving force, an optical or solid lattice, or else. Regardless, the resulting behavior of the system coupled to its environment can drastically differ from that of the isolated system. At one end of the spectrum, if the environment evolves on time scales much larger than that of the system, it is said to be frozen. In the case where the environment settles in a random frozen state, it can be accurately modeled by quenched disorder. This is the case of the celebrated Anderson model \cite{Anderson_localization} where a particle gets localized due to the presence of a random potential, leading to vanishing transport properties. The other end of the spectrum corresponds to environments with very rapid dynamics that can safely be considered Markovian and for which the Lindblad master equation has proven very useful \cite{lindbladian_spin_bath}. In between these two time scales are slowly varying baths for which a successful quantum formulation was first established by Caldeira, Leggett et al. \cite{Caldeira_Leggett_tunnelling,Bray_Moore}. They studied a single degree of freedom (spin or particle) coupled to a bath, such as the spin-boson model \cite{Caldeira_Leggett_spin_boson} or a quantum Brownian particle in a periodic potential \cite{Fisher_qbm,Schmidt_qbm}. In these models, the system is coupled to a bath of phonons that is simple enough to be traced over and recover effective dynamics for the system of interest.

More recently, interest has started shifting towards many-body versions of these models. Going from one particle to many opens the door to a new zoology of phenomena, but some single-particle effects can also persist. For instance, adding interactions to the Anderson model is conjectured to preserve localization at finite temperature through the so-called Many-Body Localization (MBL) \cite{MBL_Nandkishore,MBL_1997,MBL_RevMod}, i.e. the fact that some isolated many-body systems can fail to reach equilibrium and retain a memory of their initial conditions.

\begin{figure}[h!]
    \centering
    \includegraphics[width=7.5cm]{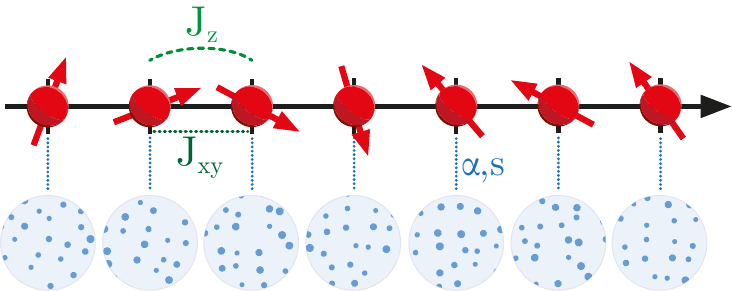}
    \caption{Schematic representation of the model. An XXZ spin chain with parameters $J_{\rm z},J_{\rm xy}$ has its spins coupled to independent and identical collections of harmonic oscillators. The spin-bath interaction is assumed to be fully described by the parameters $\alpha$ and $s$.}
    \label{fig:XXZ_spin_chain_drawing}
\end{figure}

In this work, we study the well-known XXZ spin chain coupled to local baths of harmonic oscillators (see fig.~\ref{fig:XXZ_spin_chain_drawing}). To a certain extent, this model can be viewed as a many-body extension in (1+1)D of the spin-boson model in (0+1)D \cite{Caldeira_Leggett_spin_boson,De_Filippis_2020}, or an extension of the most studied many-body localization setup, an XXZ spin chain in a random magnetic field \cite{MBL_Nandkishore}, where the fields are now replaced by local baths. The effect of the bath on its corresponding spin can be captured by an exponent $s$ (see the next section for a more formal definition) which allows to model a large variety of baths. For a spin chain at finite magnetization and any $s$, it has been shown that the bath induces fractional excitations that create a dissipative phase presenting signatures of localization \cite{Sap_ohmic,Sap_subohmic,Cazalilla_incommensurate,Ohmic_spin_boson_num}. At zero magnetization, the only known results are for $s=1$, i.e. an ohmic bath, for which a superfluid to insulator transition has been proposed in \cite{Cazalilla_commensurate,Ohmic_spin_boson_num}. In this article, we broaden the scope of these studies by focusing on the zero magnetization case and allowing for a generic exponent $s$. Our results show the existence of a Berezinsky--Kosterlitz--Thouless (BKT) phase transition between a Luttinger liquid (LL) and an antiferromagnetic phase (AFM). The exact location of the transition point depends on the bath exponent $s$ and is shifted compared to the non-dissipative spin chain for $s<1$.

The manuscript is organized as follows: section~\ref{sec:model} introduces the model and maps it to a bosonic effective field theory. Section~\ref{sec:main_results} then presents a clear overview of the main results in terms of this effective field-theoretic description. These results were derived using thorough numerical and analytical approaches. On the analytical side, a perturbative renormalization group study is presented in section~\ref{sec:pert_RG} and is complemented by a variational approach in section~\ref{sec:var_method}. These analytical predictions are then tested for several observables against an exact numerical simulation of the field theory in section~\ref{sec:num_res}. Finally, a brief discussion of the results and concluding remarks are made in section~\ref{sec:conclusion}. Unless specified otherwise (as in section~\ref{sec:var_method}), the zero temperature ($\beta \to \infty$) and thermodynamic ($L\to \infty$) limits will always be understood.

\section{Model}\label{sec:model}
The XXZ spin chain is a 1D periodic chain of N spins, total length $L$, and lattice spacing $a=\frac{L}{N}$. It evolves according to the Hamiltonian
\begin{equation}
    H_{\rm S}=\sum_{j=1}^N J_{\rm z} S_j^z S_{j+1}^z-J_{\rm xy}\left( S_j^x S_{j+1}^x+S_j^y S_{j+1}^y\right),
\end{equation}
where $S^\mu_i$ are the spin-1/2 operators. Each spin is then coupled to a set of quantum harmonic oscillators by
\begin{equation}
    H_{\rm SB}=\sum_{j=1}^N S_j^z \sum_k \lambda_k X_{k,j},
\end{equation}
and the oscillators' Hamiltonian is
\begin{equation}
    H_{\rm B}=\sum_{j=1}^N\sum_k \frac{P_{k,j}^2}{2m_k}+\frac{1}{2}m_k\Omega_k^2X_{k,j}^2.
\end{equation}
Hence the total Hamiltonian of the dissipative system $H=H_{\rm S}+H_{\rm SB}+H_{\rm B}$. Note that the baths are local, i.e. each bath has its independent spin that it acts upon and the baths are not spatially correlated. Thus, its effect can be thought of as an instance of annealed disorder $h_j(t)=\sum_k \lambda_k X_{k,j}$, acting locally through $H_{\rm SB}=\sum_{j=1}^{N} h_j(t) S_j^z$. The quenched limit of this problem $h_j(t) = h_j$ has already been studied in \cite{Giamarchi_Schulz_1987,Giamarchi_Schulz_1988,Numerical_disorder_XXZ} where a zero temperature delocalization-localization transition was found. Other bath-spin couplings are also possible and have been studied in, for example, \cite{bathengineering,Dimer_bath_coupling}. 

A common tool to study such many-body systems is the Lindblad master equation \cite{lindbladian_spin_bath} which makes use of a Markovian approximation of the bath's dynamics. Nevertheless, in this work, we will not use it and resort to an approach similar to that of Caldeira and Leggett \cite{Caldeira_Leggett_spin_boson}, which takes the bath into account in an exact manner. Following the idea of \cite{Caldeira_Leggett_spin_boson}, the bath characteristics are completely encoded in the spectral function
\begin{equation}
    \label{eq:spectral_function}
    J(\Omega)=\frac{\pi}{2}\sum_k \frac{\lambda_k^2}{\Omega_k m_k}\delta(\Omega-\Omega_k),
\end{equation}
which we assume to have the low-energy behavior
\begin{equation}
    J(\Omega)=\alpha \frac{\pi \Omega_D^{1-s}}{\Gamma(1+s)} \Omega^s \text{ for } \Omega \in [0,\Omega_D],
\end{equation}
where $s$ is the bath exponent and $\alpha$ measures the strength of the coupling to the bath. Following the classification from Caldeira and Leggett, we call $s=1$ an ohmic bath, while $0<s<1$ is a subohmic bath and $s>1$ is a superohmic bath. The rest of this section will be dedicated to deriving an effective action for our system using the bosonization technique.

\subsection{Bosonization}
To implement the bosonization procedure described in \cite{Giamarchi,von_Delft_bosonization,Sap_ohmic}, one must first map the spin chain to a one-dimensional spinless fermionic system using the Jordan--Wigner transformation. With the usual notations $c_j,c_j^\dagger$ for the ladder operators and $n_j=c_j^\dagger c_j$, the XXZ spin chain Hamiltonian becomes $H_{\rm S}=H_{\rm xy}+H_{\rm z}$, where 
\begin{align}
    H_{\rm xy}&=-\frac{J_{\rm xy}}{2}\sum_{j=1}^N\left(c^\dagger_{j+1} c_j+c^\dagger_j c_{j+1}\right),\\
    H_{\rm z}&=\sum_{j=1}^N J_{\rm z}\left(n_j-\frac{1}{2}\right)\left(n_{j+1}-\frac{1}{2}\right).
\end{align}
The hopping Hamiltonian $H_{\rm xy}$ can be diagonalized in Fourier space as $H_{\rm xy}=\sum_k \varepsilon_k c^\dagger_k c_k$ with $\varepsilon_k=-J_{\rm xy}\cos (ka)$. In the rest of the work, we will consider the spin chain to be at zero magnetization, which corresponds to a half-filled fermionic chain and a Fermi momentum $k_F=\frac{\pi}{2a}$ that is commensurate with the lattice spacing $a$. The application of the bosonization technique then requires linearizing the spectrum $\varepsilon_k$ around $k_F$ which leads to a field-theoretic description of the Hamiltonian $H_{\rm S}$ with the action $S_{\rm S}=S_{\rm LL}+S_g$, given by
\begin{eqnarray}
    S_{\rm LL} &=& \int\frac{dx d\tau}{2\pi K} \left[u\left(\partial_x\phi(x,\tau) \right)^2+\frac{1}{u}\left( \partial_\tau \phi(x,\tau)\right)^2 \right] \; \; \; \\
    S_{\rm g} &=& -\frac{gu}{2\pi^2 a^2}\int dx d\tau\cos[4\phi(x,\tau)],
\end{eqnarray}
where $x\in[0,L]$ denotes the spatial coordinate, $\tau\in[0,\beta]$ is the imaginary time coordinate with $\beta$ the inverse temperature of the system, and $S_{\rm LL}$ describes a Luttinger liquid (LL). The parameters $u,K$ are the so-called Luttinger parameters and, with $g$, are related to the microscopic parameters. The Bethe ansatz gives their exact expressions for the spin chain without dissipation (e.g. $K^{-1}_{\rm Bethe}=\frac{2}{\pi}\arccos ( -J_{\rm z}/J_{\rm xy})$) while the bosonization yields approximate expressions valid in the $J_{\rm z}\ll J_{\rm xy}$ limit (e.g. $K^{-1}_{\rm bosonization}=\sqrt{1+\frac{4J_{\rm z}}{\pi J_{\rm xy}}}$) (see appendix~\ref{appendix:bosonized_quantities} for more details). On top of this description of the isolated spin chain, one needs to add the effect of the bath captured by $H_{\rm SB}+H_{\rm B}$. This requires the bosonized expression of $S_j^z$
\begin{equation}
    \label{eq:Sz_bosonized}
    S^z_j=-\frac{a}{\pi}\nabla\phi(x_j)+\frac{(-1)^j}{\pi}\cos[2\phi(x_j)], \: x_j=ja.
\end{equation}
Using this expression to write a path integral representation of $H_{\rm B}+H_{\rm SB}$, one realizes that the bath degrees of freedom are quadratic and can therefore be integrated out. This leads to the following additional term in the action
\begin{align}
\label{eq:Salpha}
    S_\alpha=&-\frac{a}{2\pi^2}\int dx d\tau d\tau'\nonumber\\
    &\times\hspace{-0.1cm}\left[ \partial_x \phi(x,\tau)-\frac{(-1)^{x/a}}{a}\cos[2 \phi(x,\tau)]\right]\mathcal{D}(\tau,\tau')\nonumber \\
    & \times \hspace{-0.1cm}\left[ \partial_x \phi(x,\tau')-\frac{(-1)^{x/a}}{a}\cos[2 \phi(x,\tau')]\right],
\end{align}
where one has to keep in mind the underlying lattice coordinates $x_j=j a$ to make sense of the term $(-1)^{x/a}$ which arises from the commensurability of the excitation wavelength with the lattice spacing. One can show using the spectral function $J(\Omega)$ that the dissipative kernel is  $\mathcal{D}(\tau,\tau')\sim \alpha \tau_c^{s-1}|\tau-\tau'|^{-1-s}$ for $\tau > \tau_c=1/\Omega_D$ (see appendix \ref{appendix:bath_kernel} for the detailed computation). For the sake of simplicity, we relate the imaginary-time and space cutoffs as $a=u\tau_c$ which doesn't change the underlying physics. Eq.~\ref{eq:Salpha} can be further simplified by dropping the rapidly fluctuating terms $(-1)^{x/a}\partial_x \phi(x,\tau)\cos[2 \phi(x,\tau')]$, and the cross gradient term $\partial_x \phi(x,\tau)\partial_x \phi(x,\tau')$ which is irrelevant by power counting. The total bosonized action $S=S_{\rm LL}+S_g+S_\alpha$ is thus
\begin{align}
    \label{eq:bosonized_action}
    S=&\int \frac{dx d\tau}{2\pi K} \left[u\left(\partial_x\phi(x,\tau) \right)^2 +\frac{1}{u}\left( \partial_\tau \phi(x,\tau)\right)^2\right]\nonumber\\
    &-\frac{g }{2\pi^2}\int \frac{dx d\tau}{a \tau_c} \cos[4\phi(x,\tau)]\\
    &-\frac{\alpha }{2\pi^2 }\underset{|\tau-\tau'|>\tau_c}{\int \frac{dx d\tau d\tau'}{a\tau_c^{1-s}}}\frac{\cos[2 \phi(x,\tau)]\cos[2 \phi(x,\tau')]}{|\tau-\tau'|^{1+s}}.\nonumber
\end{align}
This effective action is that of a Luttinger liquid (LL) with two types of interactions: a local sine-Gordon interaction controlled by $g$, and a long-range interaction controlled by $\alpha$ which is commonly encountered in generalized XY models \cite{Giachetti_2021,Giachetti_2022} or dissipative quantum systems \cite{ribeiro2023dissipationinduced,weber2023quantum,Long_range_dissipation}. Note that the long-range interaction involving $\cos[2\phi(x,\tau)]\cos[2\phi(x,\tau')]$ can be rewritten as the sum of $\cos[2(\phi(x,\tau)+\phi(x,\tau'))]$ and $\cos[2(\phi(x,\tau)-\phi(x,\tau'))]$. In the case of an incommensurate XXZ spin chain \cite{Sap_ohmic,Sap_subohmic,Cazalilla_incommensurate}, only the latter term remains and the sine-Gordon term vanishes. It has been shown that this term gives rise to fractional excitations $|\omega_n|^s$ in the spectrum of the incommensurate spin chain. However, from our variational analysis in section \ref{sec:var_method}, we will demonstrate that this fractional term is absent from our commensurate model and is replaced by a gap. 

\section{Main results}\label{sec:main_results}
\begin{figure}[h!]
    \centering
    \includegraphics[width=7.5cm]{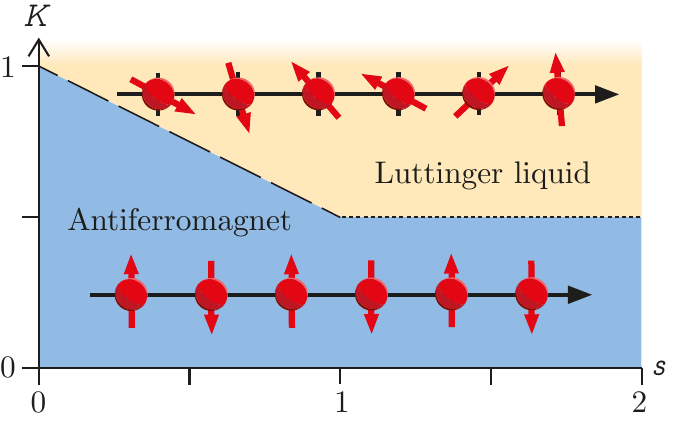}
    \caption{Phase diagram in the $(K,s)$ plane as described by the RG analysis in sec.~\ref{sec:pert_RG} and the variational method in sec.~\ref{sec:var_method} for infinitesimal value of $\alpha$ and $g$. The golden-colored phase is a gapless LL, and the blue-colored phase is a gapped AFM phase. Along the dashed line, the phase transition is controlled by the coupling $\alpha$ to the bath, while along the dotted line, it is governed by $g$ which describes the internal interactions of the spin chain.}
    \label{fig:phase_diagram}
\end{figure}

From the bosonized action in Eq.~\ref{eq:bosonized_action}, the analytical and numerical tools used in the following sections infer the zero temperature ($\beta \to \infty$) and thermodynamic ($L\to \infty$) phase diagram depicted in fig.~\ref{fig:phase_diagram}. A Luttinger liquid (LL) and an antiferromagnet (AFM) are separated by a transition that depends on the bath exponent $s$.
\begin{itemize}
    \item For a superohmic bath ($s>1$), a standard BKT transition \footnote{Here and in the following, we call BKT transition a transition whose RG equations are of the BKT type.} occurs at $K_c=1/2$ (in the limit of infinitesimal $\alpha,g$) and is driven by the coupling $g$ which comes from the internal interactions of the spin chain. The spin chain is an LL for $K \geq K_c$ while it becomes an AFM at $K<K_c$.
    \item For an ohmic bath ($s=1$), the system undergoes a BKT-like transition at $K_c=1/2$ (in the limit of infinitesimal $\alpha,g$) driven by both the coupling $g$, and the system-bath coupling denoted by the $\alpha$.
    \item For a subohmic bath ($s<1$), the coupling $\alpha$, shifts the BKT-like transition to $K_c=1-s/2$ (in the limit of infinitesimal $\alpha,g$). This increased value of $K_c$ extends the area of the AFM.
\end{itemize}
In all cases, the critical point $K_c$ increases with $\alpha$ and $g$. The two phases are characterized as follows. On the one hand, the LL is a critical phase exhibiting quasi-long range order as seen by the power-law decay of spin-spin correlation functions \cite{Giamarchi}. On the other hand, the AFM is a gapped (or massive) phase exhibiting long-range order. Although suggested by \cite{Cazalilla_commensurate}, we do not find any fractional excitations in the AFM phase, neither from the analytical study nor from the exact numerical simulations. This AFM is of the same nature as the AFM in the sine-Gordon model. The transition admits the (infinite order) order parameter $\langle \cos(2\phi)\rangle$ which is related to the antiferromagnetic spin density wave as $\langle S_j^z\rangle= \frac{(-1)^j}{\pi} \langle \cos(2\phi)\rangle$. We show that $\langle\cos(2\phi)\rangle$ vanishes with the gap of the AFM phase, thus making it an order parameter of the AFM-LL transition.

\section{Perturbative renormalization group approach}\label{sec:pert_RG}
Looking at the bosonized action in Eq.~(\ref{eq:bosonized_action}), we expect that, at large distances and times, the model presents at least two phases: a Luttinger liquid (LL) phase where both couplings $\alpha$ and $g$ are irrelevant, and a strongly interacting one where the couplings are relevant. To capture the precise location of the departure from the LL, we implement a perturbative renormalization group (RG) analysis. It turns out it is enough to compute the RG equations up to $\mathcal{O}(\alpha,g^2,\alpha g)$ to grasp the interesting physics at play. The perturbative RG analysis was done using the operator product expansion formalism \cite{Cazalilla_commensurate,cardy_RG} to respect the real-space sharp cutoff $a$ appearing in the action. The detailed computation can be found in appendix \ref{appendix:RG_OPE} and leads to the following RG equations where the dependence on the renormalization time $l$ has been made explicit,
\begin{align}
    \frac{d}{dl}&\frac{u(l)}{K(l)}=\frac{g(l)^2u(l)}{\pi^2},\label{eq:RG_eq_1}\\
    \frac{d}{dl}&\frac{1}{u(l)K(l)}=\frac{2\alpha(l)}{\pi u(l)}+ \frac{g(l)^2}{\pi^2 u(l)},\label{eq:RG_eq_2}\\
    \frac{d}{dl}&g(l)=(2-4K(l))g(l)+\alpha(l),\label{eq:RG_eq_3}\\
    \frac{d}{dl}&\alpha(l)=(2-s-2K(l))\alpha(l)+\frac{g(l)\alpha(l)}{\pi},\label{eq:RG_eq_4}
\end{align}
which agree with and extend \cite{Cazalilla_commensurate,Sap_ohmic} to generic $s$. Notice how the presence of $\alpha$ in the RG equation for $g$ implies that the bath will generate the coupling $g$ even if its microscopic (bare) value is 0. The symmetric process, i.e. $g$ generates $\alpha$, is not possible as $g$ can only enhance a non-zero $\alpha$ as seen in Eq.~\ref{eq:RG_eq_4}. When considering the effect of the bath (i.e. $\alpha\ne0$), this means that there are not two distinct phase transitions at $K_c=1/2$ and $K_c=1-s/2$ as seen from the scaling dimensions of $\alpha$ and $g$, but rather a unique transition at $K_c=\max(1-s/2,1/2)$. The corresponding phase diagram is depicted in fig.~\ref{fig:phase_diagram} and shows the existence of a LL for $K>K_c$, with $K_c=1-s/2$ for a subohmic bath ($s<1$) and $K_c=1/2$ for a superohmic bath ($s>1$). Note that the long-range cosine potential is irrelevant by power counting for $s > 2$. In this work, we are interested in understanding the effect of the bath on the system, thus we constrain the value of $s \in [0,2]$ in fig.~\ref{fig:phase_diagram}.

\subsection{Superohmic bath ($s>1$)}
\begin{figure*}[t!]
    \centering
    \includegraphics[width=1\linewidth, clip]{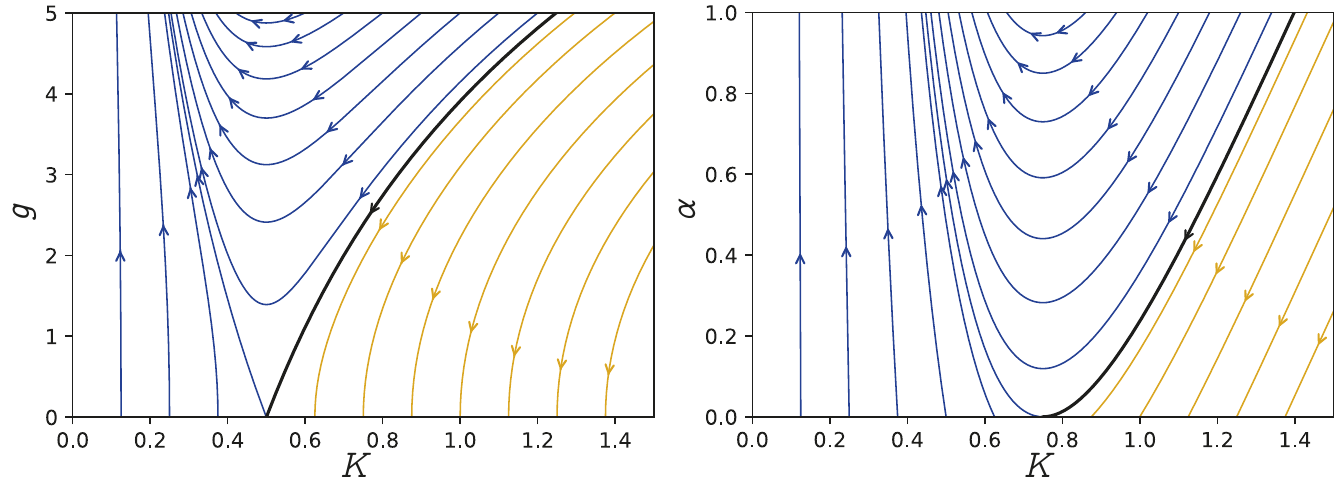}
    \caption{(left) Superohmic transition: RG flow of the couplings $g$ and $K$ according to Eqs.~\ref{eq:RG_sG1},\ref{eq:RG_sG2}. We distinguish two phases: a Luttinger liquid (LL) depicted in gold, and an antiferromagnetic phase (AFM) in blue. The critical point is at $K_c=1/2$. (right) Subohmic transition (here $s=0.5$): RG flow of the couplings $\alpha$ and $K$ according to Eqs.~\ref{eq:RG_com1},\ref{eq:RG_com2}. We distinguish two phases: a Luttinger liquid (LL) in gold, and an antiferromagnetic phase (AFM) in blue. The critical point is at $K_c=1-s/2=0.75$.}
    \label{fig:RG_flow_all}
\end{figure*}

For a superohmic bath ($s>1$), near the transition point $K_c=1/2$, the coupling $\alpha$ has a scaling dimension $2-s-2K_c=1-s<0$ which is strongly irrelevant. It can therefore be safely ignored and the RG equations become
\begin{align}
    \frac{d}{dl}&u=0,\\
    \frac{d}{dl}&\frac{1}{K}=\frac{g^2}{\pi^2},\label{eq:RG_sG1}\\
    \frac{d}{dl}&g=(2-4K)g,\label{eq:RG_sG2}
\end{align}
which are the one-loop RG equations of the sine-Gordon model \cite{NdupuisCMUG2,Giamarchi}. The associated RG flow is shown in fig.~\ref{fig:RG_flow_all} and is known to belong to the BKT universality class. According to the standard BKT phenomenology, we expect a gap $\Delta$ to appear for $K>K_c$. A well-known result derived from the RG equations is that $\Delta \sim \exp (-C(g-g_c)^{-p})$ near the the transition, with $p=1/2$ \cite{Giamarchi}.

\subsection{Ohmic bath ($s=1$)}
The ohmic bath ($s=1$) is the most studied type of bath in the literature \cite{Sap_ohmic,Cazalilla_commensurate,Ohmic_spin_boson_num}. In this setting, the scaling dimensions of $\alpha$ and $g$ both vanish at $K_c=1/2$, which means the transition is driven by both couplings simultaneously. The RG equations Eqs.~\ref{eq:RG_eq_1},\ref{eq:RG_eq_2},\ref{eq:RG_eq_3},\ref{eq:RG_eq_4} cannot be simplified any further (as is the case for superohmic or subohmic baths). This leads to an RG flow similar to that of the BKT transition, the main difference being that it is governed by two couplings ($\alpha$ and $g$) instead of one. 

We have shown in appendix~\ref{appendix:p_ohmic} that this BKT-like transition is characterised by a gap closing as $\Delta \sim \exp (-C(\alpha-\alpha_c)^{-p})$ (or, equivalently, $\exp (-C'(g-g_c)^{-p})$ if one tunes the coupling $g$ instead $\alpha$) near the transition with $p=\frac{\sqrt{79}+3}{35}\simeq0.3396$. BKT-like transitions with a parameter $p\ne 1/2$ have been previously found, for example, in the case of 2D melting \cite{Nelson1978} and long-range Ising models \cite{Cardy1981}. However, to the best of our knowledge, this particular value of $p$ has not yet been reported in the literature.

\subsection{Subohmic bath ($s<1$)}
\label{subsec:RG_subohmic}
For a subohmic bath ($s<1$), the coupling $g$ is strongly irrelevant near the transition point $K_c=1-s/2$ since its scaling dimension is $2-4K_c=2s-2<0$. The RG equations can therefore be simplified by discarding $g$ to give
\begin{align}
    \frac{d}{dl}&\frac{u}{K}=0,\\
    \frac{d}{dl}&\frac{1}{uK}=\frac{2\alpha}{\pi u},\label{eq:RG_uK}\\
    \frac{d}{dl}&\alpha=(2-s-2K)\alpha.\label{eq:RG_alpha}
\end{align}
The LL action appears to be renormalized along the imaginary-time direction ($\frac{d}{dl}\frac{1}{uK}\ne 0$) but not along the space direction ($\frac{d}{dl}\frac{u}{K}=0$)\footnote{Pushing the RG equations to higher orders shows that $\frac{d}{dl}\frac{u}{K}=\mathcal{O}(\alpha^2)$. This is to be contrasted with the incommensurate spin chain \cite{Sap_ohmic,Sap_subohmic} for which $\frac{d}{dl}\frac{u}{K}=0$ is non-perturbative. This is because the system-bath interaction in the incommensurate spin chain action is invariant under the statistical tilt symmetry $\phi(x,\tau)\to\phi(x,\tau)+ \mu x$, where $\mu$ is a constant.}. This anisotropy is due to the long-range interaction in Eq.~\ref{eq:bosonized_action} which only spans the imaginary-time direction. Since the ratio $\frac{u}{K}$ is constant, Eqs.~\ref{eq:RG_uK},\ref{eq:RG_alpha} can be written in terms of $K$ and $\sqrt{\alpha}$ only
\begin{align}
    \frac{d}{dl}&\frac{1}{K}=\frac{\sqrt{\alpha}^2}{\pi},\label{eq:RG_com1}\\
    \frac{d}{dl}&\sqrt{\alpha}=(1-s/2-K)\sqrt{\alpha},\label{eq:RG_com2}
\end{align}
which are exactly the same type of equations as Eqs.~\ref{eq:RG_sG1},\ref{eq:RG_sG2} describing the sine-Gordon model. This shows that this phase transition is also BKT-like and doesn't belong to a totally new universality class as suggested by \cite{1Dquantum_new_phase_tran}. The RG flow associated to these equations is depicted in fig.~\ref{fig:RG_flow_all}. Because of the identification with the subohmic RG equations, the gap closes as  $\Delta \sim \exp (-C(\alpha-\alpha_c)^{-p})$ near the the transition with $p=1/2$.

\section{Variational method}\label{sec:var_method}
While the perturbative RG is tailored to capture the end of the Luttinger liquid (LL), it fails at describing correctly the physics of the antiferromagnetic phase (AFM). This is why we resort to a variational method {\it à la} Feynman \cite{feynman_stat_mech,Giamarchi} to capture the essential properties of the bulk of the phases. This technique aims at finding an approximate action $S_{\rm var}$ that is very similar to $S$ while being simple enough to allow for analytical computations. We consider a quadratic ansatz
\begin{equation}
    S_{\rm var}=\frac{1}{2}\int \frac{dq}{2\pi}\frac{d\omega_n}{2\pi} \phi^\star(q,\omega_n) G^{-1}_{\rm var} (q,\omega_n) \phi(q,\omega_n),
\end{equation}
where $G^{-1}_{\rm var} (q,\omega_n)=1/G_{\rm var}(q,\omega_n)$ is to be determined, $\phi(q,\omega_n)$ is the Fourier transform of $\phi(x,\tau)$ given by $\phi(x,\tau) = \int \frac{dq}{2\pi}\frac{d\omega_n}{2\pi} \phi(q,\omega_n) e^{i\left(qx-\omega_n \tau \right)}$ with $\omega_n=\frac{2 \pi n}{\beta}$ and $q=\frac{2\pi n}{L}$ the bosonic Matsubara frequencies and the momenta of the system, respectively. The distance from the original action to the variational action is then defined through the variational free energy $F_{\rm var}=  -T \log Z_{\rm var} + T \langle S-S_{\rm var}\rangle_{\rm var}$ ($\langle \ldots \rangle_{\rm var}$ stands for the average with respect to $S_{\rm var}$). Indeed, the true free energy of the original field theory $F=-T \log Z=T \log Z_{\rm var} - T \log \langle e^{-(S-S_{var})}\rangle_{\rm var}$, is always upper bounded by $F_{\rm var}$ due to the convexity of the exponential. This defines the best variational action as that which minimizes $F_{\rm var}$. The goal is thus to solve the vanishing gradient condition $\frac{\delta F_{\rm var}}{\delta G^{-1}_{\rm var} (q,\omega_n)}=0$. It is possible to give an explicit expression for $F_{\rm var}$ which in turn yields the vanishing gradient equation
\begin{widetext}
\begin{align}
    \label{eq:selfconsist}
    G^{-1}_{\rm var}(q,\omega_n)=&\frac{1}{\pi K}\left[ uq^2+\frac{\omega_n^2}{u}\right]+\frac{8g}{\pi^2a \tau_c}\exp \left[-\frac{2}{\pi^2}\int dq' d\omega_n' G_{\rm var}(q',\omega_n')\right]\\
    &+\frac{2\alpha }{\pi^2a \tau_c^{1-s}}\int_{\tau_c}^\infty \frac{d\tau}{\tau^{1+s}}\sum_{\varepsilon=\pm 1}(1+\varepsilon \cos \omega_n \tau )\exp \left[-\frac{1}{\pi^2}\int dq' d\omega_n' (1+\varepsilon \cos \omega_n' \tau ) G_{\rm var}(q',\omega_n')\right].\nonumber
\end{align}  
\end{widetext}

\subsection{Phase diagram at infinitesimal coupling}
In the absence of interactions, i.e. $\alpha=g=0$, Eq.~\ref{eq:selfconsist} reduces to $G^{-1}_{\rm LL}(q,\omega_n)=\frac{1}{\pi K}\left[ \frac{\omega_n^2}{u}+uq^2\right]$ which is just the propagator of the Luttinger liquid. When reintroducing the interaction terms, we expect a gap to appear, like in the sine-Gordon model \cite{Giamarchi}, and maybe some fractional excitations, as in \cite{Cazalilla_commensurate}, so we postulate the following ansatz for the propagator in the AFM
\begin{align}
    G^{-1}_{\rm AFM}(q,\omega_n)=&\frac{1}{\pi K}\left[uq^2+ \frac{\omega_n^2}{u}+\nu |\omega_n|^s+\frac{\Delta^2}{u}\right],
    \label{eq:self_con_var}
\end{align}
where the parameters $\Delta$ and $\nu$ are to be determined. This ansatz is expected to be valid in the $\alpha,g \to 0$ limit where one can neglect the renormalization of the Luttinger parameters $u,K$ into $u_r,K_r$ (more on this in the next subsection). The first observation we make is that $\nu=0$. This can be understood from the fact that the $|\omega_n|^s$ excitations can only come from the bath-dependent $\alpha$ term in Eq. \ref{eq:self_con_var}, and the $\varepsilon=1$ contribution exactly cancels out that of $\varepsilon=-1$. It is worth noticing that unlike \cite{Cazalilla_commensurate}, where a similar action with an ohmic bath was studied, this means that we do not find any fractional excitation $|\omega_n|^s$ in our system for any $s$. We now move on to the determination of the gap $\Delta$. Setting $q=0$, $\omega_n=0$ in Eq.~\ref{eq:selfconsist} yields an equation for $\Delta$ that can be written as (see appendix \ref{appendix:gap_eq} for the detailed computation)
\begin{align}
    \label{eq:delta_phase_trans} 
    \frac{(\Delta\tau_c)^{2-4K}}{K}=&g\frac{8}{\pi}+\alpha\frac{4 e^{2K\gamma_E}}{\pi}\frac{(\Delta\tau_c)^{s-2K}-1}{2K-s},
\end{align}
where $\tau_c$ is the imaginary-time UV cutoff, and $\gamma_E$ is Euler's gamma constant. As usual with this variational approach, Eq.~\ref{eq:delta_phase_trans} is valid deep in the AFM phase where $g,\alpha \ll K_c-K$ \cite{Giamarchi} and in the limit of small $\Delta \tau_c$. To arrive at the phase diagram shown in fig.~\ref{fig:phase_diagram}, we write Eq.~\ref{eq:delta_phase_trans} as $\Delta^{2-4K}= a_1 + a_2 \frac{\Delta^{s-2K}}{2K-s}$ with $a_1,a_2$ two constants independent of $\Delta$. As we approach the transition $K\to K_c^-$, the gap $\Delta$ is expected to vanish, so we retain only the leading terms in the previous gap equation. This suggests to distinguish three cases depending on the sign of $2K_c-s$.
\begin{figure*}[t!]
    \centering
    \includegraphics[width=1\linewidth, clip, trim=80 160 120 120]{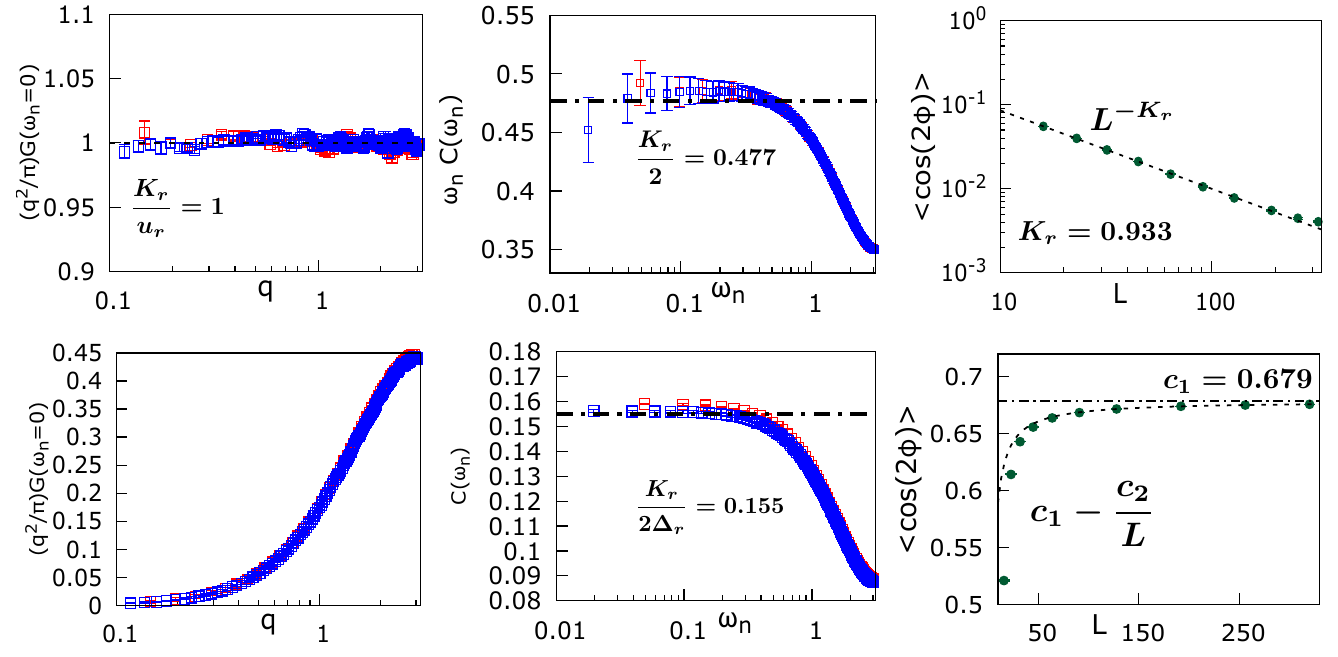}
    \caption{Numerical evaluation of observables for the commensurate dissipative spin chain. The values of the parameters are taken to be $s=0.5$, $K=1$, $g=0$, $u=1$, and $\beta$ is scaled as $L$. For the \textit{left} and \textit{middle} column, red and blue colors denote $L=128$ and $L=320$ respectively. All the quantities have been averaged over 10000-20000 configurations and the value of time-step $dt$ is $0.5$. (top) and (bottom) rows correspond to $\alpha=1$ (LL) and $\alpha=5$ (AFM), respectively. (left) Susceptibility $\chi$ as a function $q$. For $\alpha=1$, it stays finite and constant whereas for $\alpha=5$, it vanishes at small $q$. (middle) The behavior of $C(\omega_n)$ (Eq. \ref{eq:Cw}) as a function of $\omega_n$. For $\alpha=1$, $\omega_n C(\omega_n)$ saturates to a constant $K_r/2=0.477$; for $\alpha=5$, $C(\omega_n)$ becomes a constant $K_r/(2\Delta_r)=0.155$. (right) Behavior of the order parameter $\langle \cos \left[2\left(\phi-\phi_{\rm CoM} \right)\right] \rangle$  as a function of system size $L$. For $\alpha=1$, the order parameter decays as $L^{-K_r}$ with $K_r = 0.933$, whereas for $\alpha=5$, it increases and saturates to a constant as $c_1 - c_2/L$ with $c_1 = 0.679$ and $c_2=0.894$.}
    \label{fig:result}
\end{figure*}
\begin{itemize}
   \item $s-2K_c>0$: The gap equation reduces to $\Delta= a_1^{\frac{1}{2-4K}}$ which tells us that the critical point is $K_c = 1/2$ with $K<K_c$ corresponding to the gapped phase, while $K>K_c$ is gapless. We also understand that this gapped phase solution is valid only if $s>2K_c=1$, i.e. the bath is superohmic ($s>1$).
   
   \item $s-2K_c<0$: Retaining only the leading terms yields $\Delta \sim a_2^{\frac{1}{2-s-2K}}$, signifying that $K_c = 1 - \frac{s}{2}$ and the gapped phase is for $K<K_c$. In this regime $s < 2K_c= 2-s$, which means the bath is subohmic ($s<1$).
   
    \item $s-2K_c=0$: For $K$ close to $K_c$, one expands $a^{2K-s}=1+(2K-s)\log a$ since $2K-s\ll 1$, which shows that to leading order $\Delta^{2-4K}\propto-\log (\Delta \tau_c)$. This yields $K_c = 1/2$ and thus $s=1$ which is an ohmic bath. This matches with the two previous cases but the gap is now discontinuous at the transition. This is a known artifact of the variational method \cite{Cazalilla_commensurate,Giamarchi}.
\end{itemize}
In the spin chain picture, this gapped phase turns out to be an AFM. Indeed, by averaging Eq.~\ref{eq:Sz_bosonized} over all field configurations, one arrives at 
\begin{equation}
\langle S^z_j \rangle=\frac{(-1)^{\frac{x_j}{a}}}{\pi}\langle \cos[2\phi(x_j)]\rangle,
\label{eq:OP_significance}
\end{equation}
which denotes the existence of an AFM phase with amplitude $\frac{1}{\pi}\langle \cos (2\phi) \rangle$. Using the variational propagator $G^{-1}_{\rm AFM}$ for the gapped phase, one then proves that $\langle \cos (2\phi) \rangle = \left(\Delta \tau_c\right)^{K}$, which can be obtained from the $\Delta \tau_c \ll 1$ limit of Eq.~\ref{eq:gap_OP}. This implies that $\langle \cos (2\phi) \rangle$ can be used as an order parameter for this transition as this quantity vanishes in the LL and is finite in the AFM. Moreover, looking back at the gap expressions derived from the RG near the transition ($\Delta \sim \exp(-C(g-g_c)^p)$ or $\Delta \sim \exp(-C(\alpha-\alpha_c)^p))$, it appears that all derivatives of the order parameter vanish at the transition. This corroborates the scenario of a BKT-like phase transition which is known to be of infinite order. Physically, this order parameter is associated with the spontaneous symmetry breaking of the discrete shift symmetry $\phi(x,\tau)\to \phi(x,\tau)+\frac{n\pi}{2}$ of the bosonized field. Indeed, we expect $\langle \cos (2\phi) \rangle=\langle \cos (2\phi+\pi) \rangle=0$ when the symmetry holds, which shows that the AFM is the symmetry-broken phase.

The fact this order parameter is associated with an AFM phase is not surprising as in the case of a dissipative incommensurate spin chain, the ordered phase is a spin density wave of wavelength $\pi/k_F$ \cite{Sap_subohmic}. Putting $k_F = \frac{\pi}{2a}$, one recovers an AFM spin density wave with a $2a$ wavelength. All these results confirm the phase diagram obtained from the RG (see fig.~\ref{fig:phase_diagram}). There are two phases separated by $K_c=\max(1-s/2,1/2)$; for $K>K_c$, the system remains an LL, while for $K<K_c$ it becomes an AFM.

\begin{figure}
    \centering
    \includegraphics[width=1\linewidth, clip, trim=140 115 160 115]{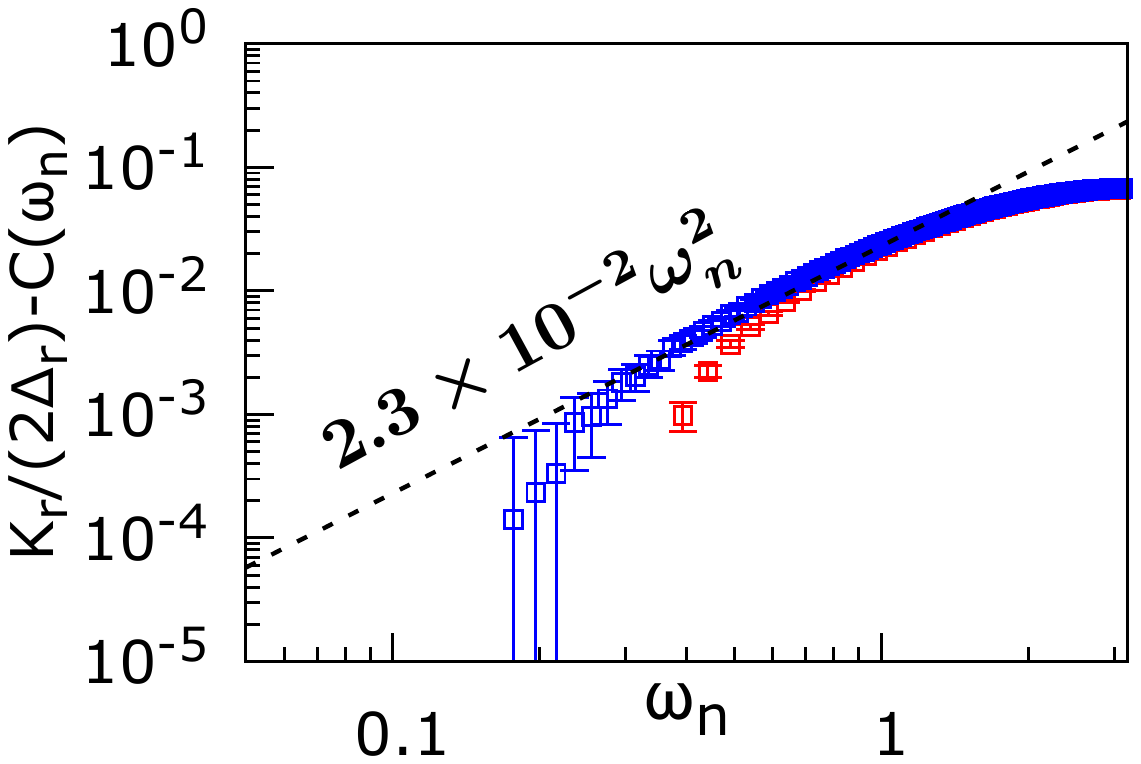}
    \caption{$K_r/(2\Delta_r)-C(\omega_n)$ for the dissipative phase ($s=0.5, K=1, g=0, \alpha=5$). This quantity fits well with $2.3 \times 10^{-2} \omega_n^2$ for small $\omega_n$, which is the sub-leading $\omega_n$ dependence of the propagator in the AFM. The red and blue colors denote two different system sizes $L=128$ and $L=320$ respectively.}
    \label{fig:gapped_no_frac}
\end{figure}

\subsection{Phase diagram at finite coupling}
In the previous subsection, we showed how the simple ansatz $G^{-1}_{\rm AFM}(q,\omega_n)=\frac{1}{\pi K}\left[uq^2+ \frac{\omega_n^2}{u}+\frac{\Delta^2}{u}\right]$ was enough to capture the correct location of the phase transition when $\alpha,g\to 0$. However, from the RG results we expect that starting at any value of $K>K_c$ and with $g$ fixed, the system remains in the LL phase for $0<\alpha<\alpha_c$ and switches to an AFM for $\alpha>\alpha_c$. At $\alpha=\alpha_c$, the system should be a LL with renormalized Luttinger parameter $K_r=K_c$. The same analysis is of course valid if tuning $g$ while keeping $\alpha$ constant. One could therefore wonder if a more general ansatz $G^{-1}_{\rm var}(q,\omega_n)=\frac{1}{\pi K_r}\left[u_rq^2+ \frac{\omega_n^2}{u_r}+\frac{\Delta_r^2}{u_r}\right]$, where $u_r$, $K_r$, $\Delta_r$ are fitting parameters to be determined, could capture the renormalization of the Luttinger parameters as predicted by the RG. It appears that this ansatz works well deep in the phase for finite $\alpha$. However, it fails to recover the lowest order RG equations close to the transition (see appendix \ref{appendix:var_vs_RG}). In the next section, we will nonetheless use the respective renormalized quantities, denoted with a subscript $r$ (e.g. the renormalized value of $K$ is $K_r$ and so on), as fitting parameters for our numerical analysis done at finite $g,\alpha$.

\section{Numerical Results}\label{sec:num_res}

In this section, we calculate different observables, both analytically and numerically, to observe the signature of the phase transition and physically characterize the ordered phase (AFM). For our numerical analysis, we simulate the Langevin dynamical equation for the field $\phi$ associated with the equilibrium probability distribution $P_{\rm eq.}[\phi]=e^{-S\left[ \phi \right]}$. The Langevin equation is $\frac{d \phi_{ij}(t)}{dt} = - \frac{\delta S\left[\phi_{ij}(t) \right]}{\delta \phi_{ij}(t)} + \eta_{ij}(t)$, where $i,j$ are the discretized indices for imaginary-time $\tau$ and space $x$ respectively, $t$ denotes the Langevin time (alternatively, the simulation time) and $\eta(t)$ is Gaussian white noise with $\langle \eta_{ij} (t) \rangle = 0$ and $\langle \eta_{ij} (t) \eta_{i'j'} (t') \rangle = 2 \delta_{i,i'}\delta_{j,j'}\delta(t-t')$. Note that the noise $\eta$ is used only for thermalizing the Langevin equation and is not to be confused with the noise coming from the dissipative bath, which has already been taken into account in the action $S[\phi]$. Using this numerical technique, we simulate configurations for fixed values of $s,K$ and for different values of $\alpha$. For each set of parameters, we extract field configurations of different sizes $L\times \beta$. We scale $L$ as $\beta$, keeping the BKT dynamic scaling  $z=1$ in mind; and on these configurations, we calculate several observables to characterize the two phases. In the following, we show the results for $s=0.5$ (subohmic bath), $K=1$, $g=0$, and $u=1$. Results for $s=1$ (ohmic bath) and $s=1.5$ (superohmic bath) can be found in appendix \ref{App:more_num_res}.

\subsection{Relevant observables}
The first quantity that we calculate is the static susceptibility of the spin chain $\chi = \lim_{q \to 0} (q/\pi)^2 G\left(q,\omega_n=0\right)$, also known as the compressibility of the field. Using the LL propagator $G_{\rm LL} = \pi K_r \left[u_r q^2 + \frac{\omega_n^2}{u_r} \right]^{-1}$ derived in section~\ref{sec:var_method}, it appears that the static susceptibility is finite in the LL phase and is given by $\chi_{\rm LL} = K_r/(u_r \pi) $. On the other hand, using $G_{\rm AFM}=\pi K_r \left[u_r q^2 + \frac{\omega_n^2}{u_r} + \frac{\Delta_r^2}{u_r}\right]^{-1}$ shows that $\chi_{\rm AFM}$ vanishes as $ \frac{K_r u _r}{\Delta_r^2} q^2$ in the gapped phase and for $q \to 0$. These analytical predictions are then compared to the data from the Langevin simulation. Numerically, the Green's function $G\left(q,\omega_n\right)$ is obtained by computing the correlation function $\langle \phi(q,\omega_n)\phi(-q,-\omega_n) \rangle = \langle |\phi(q,\omega_n)|^2 \rangle$, as the fields are real and bosonic. The associated numerical results are depicted in fig.~\ref{fig:result}, \textit{left} and support our analytical predictions. In the LL ($\alpha=1$), the quantity $\lim_{q \to 0} (q^2/\pi) G\left(q,\omega_n=0\right) = \pi \chi$ is equal to $K_r/u_r =1$ (\textit{top row}). On the other hand, in the AFM ($\alpha=5$), the $q \to 0$ limit vanishes (\textit{bottom row}).
\begin{figure*}[t!]
    \centering
    \includegraphics[width=1\linewidth, clip, trim=60 195 40 165]{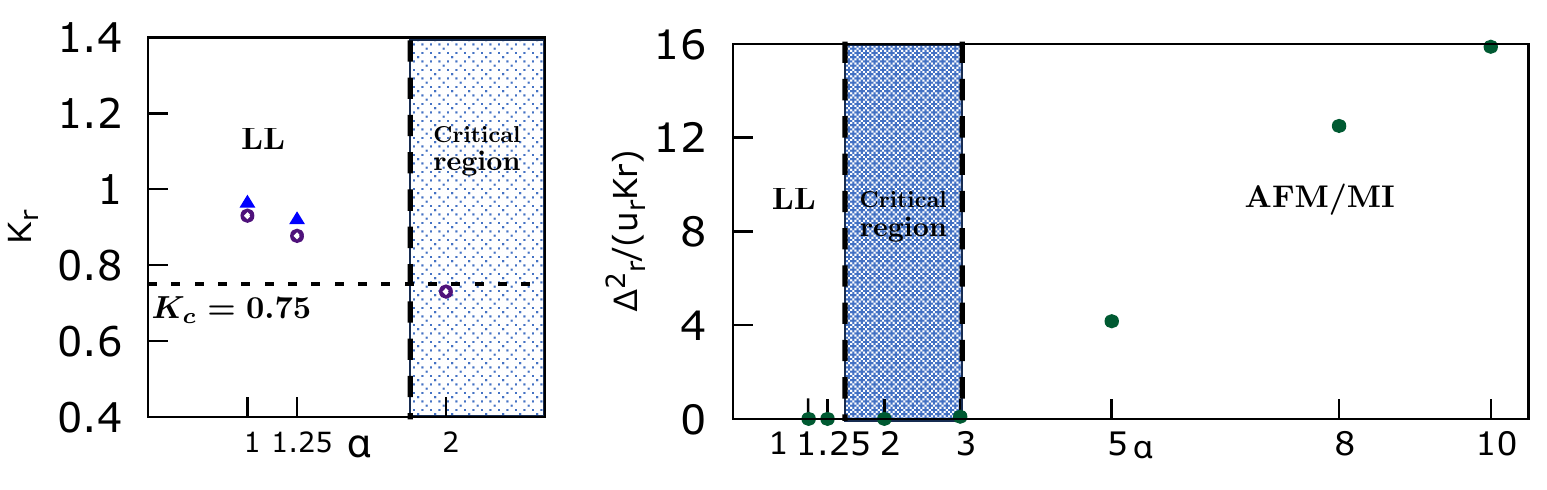}
    \caption{Numerical phase diagram obtained for $K=1,u=1,g=0,s=0.5$ by varying $\alpha$. (left) $K_r$ in the LL phase as a function of $\alpha$, extracted from the order parameter $\langle \cos\left[2\left(\phi - \phi_{\rm CoM}\right) \right] \rangle$ (purple circles) and from $C(\omega_n)$ (blue triangles). $K_r$ decreases to $K_c = 0.75$ as $\alpha$ is increased, signalling a BKT transition. (right) As $\alpha$ is increased, the gap (green dots) first becomes finite for $\alpha = 3$ and then increases.}
    \label{fig:num_phase_diag}
\end{figure*}

The next quantity that we calculate is given by $C(\omega_n) = \frac{1}{\pi L}\sum_{q=-\infty}^{\infty} G(q,\omega_n)$. In the thermodynamic limit, the sum over $q$ can be replaced with the integral $\frac{1}{2\pi} \int_{-\infty}^{\infty} dq$. In this limit, using our variational action from section~\ref{sec:var_method} we expect that for a small-$\omega_n$ limit,
\begin{equation}
   C(\omega_n \to 0) = 
   \begin{cases}
       \frac{K_r}{2 \omega_n} \ & \mbox{LL,}\\
       \frac{K_r}{2\Delta_r} \left[1 - \frac{1}{2} \left( \frac{\omega_n}{\Delta_r} \right)^2 \right]\ & \mbox{AFM.}
   \end{cases}
\label{eq:Cw}
\end{equation}
Comparing this to the numerical results in fig.~\ref{fig:result}, \textit{middle}, we see that indeed for small $\alpha$ (LL phase), $\omega_n C(\omega_n)$ goes to a constant for $\omega_n \to 0$; whereas $C(\omega_n)$ itself saturates to a constant for a higher value of $\alpha$ (AFM phase), as analytically predicted. This confirms the existence of a gap in the low-energy spectrum of the dissipative phase. One can also check the sub-leading $\omega_n$ dependence by numerically calculating $\frac{K_r}{2\Delta_r} - C(\omega_n)$. Figure \ref{fig:gapped_no_frac} shows that this term varies as $\propto \omega_n^2$, which backs up our variational prediction of the absence of a fractional laplacian term $|\omega_n|^s$.

Finally, we compute the order parameter $\langle\cos \left[ 2\left(\phi - \phi_{\rm CoM} \right) \right]\rangle$, where $\phi_{\rm CoM} = \frac{1}{\beta L} \sum_{x,\tau} \phi(x,\tau)$ is the center of mass (CoM) of the configuration. The field has been offset by its CoM to suppress the contribution $G(0,0)$ to the order parameter which would otherwise diverge in the gapless LL phase. In appendix \ref{App:OP}, we analytically show that in the LL phase (with $\beta=L$), the order parameter vanishes as $L^{-K_r}$, whereas it increases and saturates to a constant in the AFM as $c_1 - c_2/L$, where $c_1$ and $c_2$ are both positive constants. Numerical results in fig.~\ref{fig:result}, \textit{right}, corroborate this scenario. This confirms our prediction that the dissipative bath can induce a quantum phase transition on a 1D XXZ spin chain by spontaneously breaking the discrete shift symmetry $\phi \to \phi + \frac{n\pi}{2}, n \in \mathcal{Z}$, and the ordered phase is a gapped AFM phase.

\subsection{Microscopic parameters}
One can numerically extract relevant microscopic parameters in both phases from fitting the previous observables. In the LL phase, the relevant quantity is $K_r$, which can be extracted from the calculation of the order parameter by fitting it as a function of system size $L$ (recall that it should scale as $\sim L^{-K_r}$), and from $C(\omega_n)$ which saturates to the constant $K_r/(2\Delta_r)$ as $\omega_n \to 0$. In the gapped phase, the extraction of the parameters is slightly more tricky as, at the lowest order, the quantitative behavior of the phase is regulated by the gap term. By computing $\lim_{q \to 0} \chi(q)/q^2$, one can nonetheless extract the value of the inverse of the gap $u_r K_r/\Delta_r^2$. The behaviors of these parameters are given in fig.~\ref{fig:num_phase_diag}. From the plots, we see that $K_r$ starts decreasing as $\alpha$ is increased and approaches $K_c = 0.75$ at the transition, after which $\Delta_r$ becomes finite and increases as $\alpha$ is increased. This tells us that for $K=1$, $u=1$, $g=0$ and $s=0.5$, the transition happens around $\alpha_c \in (2,3)$.

\section{Discussions and conclusions}\label{sec:conclusion}
In this work, we investigated the zero temperature properties of an XXZ spin chain at zero magnetization coupled to local baths of phonons \emph{à la} Caldeira and Leggett. Using the bosonization procedure, this system was mapped onto an effective field theory (with bosonic field $\phi$) which could thus be tackled using powerful analytical and numerical field-theoretic tools. We confirmed the existence of a BKT-like phase transition between a Luttinger liquid (LL) exhibiting quasi-long-range order and an antiferromagnetic (AFM) phase with long-range order. The variational method predicts that the AFM does not present any fractional excitations as suggested by \cite{Cazalilla_commensurate}. The exact location of the phase transition depends on the type of bath studied through the bath exponent $s$. For superohmic baths ($s>1$), dissipation does not affect the transition which remains that of the isolated spin chain to the leading order. This transition is the standard BKT one. However, for subohmic baths ($s<1$), the transition is shifted and the AFM eats into the LL. Although akin to the standard BKT universality class, this transition differs from the standard one. For subohmic baths, its location is shifted, while for ohmic baths the scaling of the gap near the transition is altered (see section~\ref{sec:pert_RG}). It appears that, for all bath exponent $s$, the transition is associated with the spontaneous symmetry breaking of the order parameter $\langle \cos(2\phi)\rangle$. This quantity identifies with the amplitude of the antiferromagnetic spin density wave $S^z_j\sim(-1)^j\langle \cos(2\phi)\rangle$. As with a standard BKT transition, all derivatives of this order parameter vanish at the transition, signaling an infinite order phase transition.

While the variational method used in section \ref{sec:var_method} successfully captured the main properties of both phases, it makes use of a rather crude approximation amounting to replacing the system's highly non-linear action by a quadratic one. It is a well-established fact that such an approximate action fails at capturing topological excitations of the system such as solitons or instantons \cite{Giamarchi,rajaraman_solitons}. For a generic AFM with a vanishing linear conductivity, such excitations might lead to the restoring of non-linear terms \cite{soliton_pair_1978,dissipative_sine_gordon}. For our model there are no instantons with finite action in the zero temperature limit  (see chapter 3.2 of \cite{rajaraman_solitons} for a proof of this statement), and the system is truly locked in one of the minima of the potential. Nonetheless, one might worry about the effect of solitons, for instance on the transport properties, as for the pure sine-Gordon model. We leave for further investigation the understanding of this point. Moreover, the study of the same model at finite temperature, where instantons connecting degenerate minima play an important role, seems an interesting path to explore.

Another direction we wish to explore is that of the competition between bath-induced localization, as studied in this work and \cite{Sap_ohmic,Sap_subohmic}, and disorder-induced Anderson localization. One such example can be found in \cite{Ros} where a quantum phase transition was found between an Anderson localized phase and a Zeno localized phase in a one-dimensional non-interacting system. Finally, valuable insights could probably be gained by mapping this system to a Coulomb-like gas using the general idea of the sine-Gordon to the 2D Coulomb gas mapping \cite{RevMod_Coulomb_gas_Minnhagen,bouverotdupuis2024mapping}.

\begin{acknowledgments}
{\it Acknowledgments:} This
work was granted access to the HPC resources of IDRIS under the allocation 2022-(AD011013581R1) made by GENCI. We thank Thierry Giamarchi, Thibaud Maimbourg, Manon Michel and Xiangyu Cao for useful discussions. We are also grateful to Edmond Orignac for pointing out the role of topological solutions in our system.
\end{acknowledgments}

\appendix

\section{Field-theoretical couplings from bosonization}
\label{appendix:bosonized_quantities}
The couplings $K,u,g$ appearing in the bosonic field-theory in Eq.~\ref{eq:bosonized_action} are related to the microscopic parameters $J_{\rm z},J_{\rm xy},a$ of the XXZ spin chain. Bosonization predicts the following correspondence
\begin{align}
    K&=\sqrt{1+\frac{4J_{\rm z}}{\pi J_{\rm xy}}}^{-1},\\
    u&=aJ_{\rm xy}\sqrt{1+\frac{4J_{\rm z}}{\pi J_{\rm xy}}},\\
    g&=\frac{J_{\rm z}}{J_{\rm xy}}\sqrt{1+\frac{4J_{\rm z}}{\pi J_{\rm xy}}}^{-1},
\end{align}
which is known to be valid in the regime $J_z \ll J_{\rm xy}$. While an exact Bethe ansatz solution for the isolated XXZ spin chain exists \cite{Haldane_bethe}, such a solution does not (yet) exist for the dissipative spin chain studied in this article.

\section{Dissipative kernel}
\label{appendix:bath_kernel}
When integrating out the bath degrees of freedom, one generates the dissipative kernel $\mathcal{D}(\tau,\tau')=\sum\limits_k \mathcal{D}_k(\tau,\tau')$, where $\mathcal{D}_k^{-1}(\tau,\tau')=\frac{m_k}{\lambda_k^2}\delta(\tau-\tau')(\Omega_k^2-\partial_{\tau'}^2)$. To recover $\mathcal{D}(\tau,\tau')$, one must therefore invert $\mathcal{D}^{-1}_k(\tau,\tau')$ which amounts to finding $\mathcal{D}_k(\tau,\tau')$ such that
\begin{equation}
\label{eq:inverse_def}
    \int d\tau' \mathcal{D}_k(\tau,\tau')\mathcal{D}^{-1}_k(\tau,\tau')=\delta(\tau-\tau'').
\end{equation}
Using the following Fourier and inverse transform conventions
\begin{align}
\label{eq:operator_forward_FT}
    \mathcal{D}_k(\omega,\omega')&=\int d\tau d\tau' e^{i\omega\tau}\mathcal{D}_k(\tau,\tau')e^{-i\omega'\tau'},\\
    \label{eq:operator_backward_FT}
    \mathcal{D}_k(\tau,\tau')&=\int \frac{d\omega}{2\pi} \frac{d\omega'}{2\pi} e^{-i\omega\tau}\mathcal{D}_k(\omega,\omega')e^{i\omega'\tau'},
\end{align}
Eq.~\ref{eq:inverse_def} can be written in Fourier space as
\begin{equation}
\label{eq:inverse_def_fourier}
     \int d\omega'\mathcal{D}_k(\omega,\omega')\mathcal{D}_k^{-1}(\omega',\omega'')=4\pi^2 \delta(\omega-\omega'').
\end{equation}
Let's apply this to the kernel $\mathcal{D}_k^{-1}(\tau,\tau')$. According to Eq.~\ref{eq:operator_forward_FT}, its Fourier transform is $\mathcal{D}^{-1}_k(\omega,\omega')=\frac{m_k}{\lambda_k^2}2\pi\delta(\omega-\omega')(\Omega_k^2+{\omega'}^2)$. From this expression and Eq.~\ref{eq:inverse_def_fourier}, it is clear that $\mathcal{D}_k(\omega,\omega')=\frac{\lambda_k^2}{m_k}\frac{2\pi\delta(\omega-\omega')}{\Omega_k^2+{\omega'}^2}$.
This implies, along with Eq.~\ref{eq:operator_backward_FT}, that
\begin{align}
    \mathcal{D}_k(\tau,\tau')&=\frac{\lambda_k^2}{m_k}\int \frac{d\omega}{2\pi} \frac{e^{-i\omega(\tau-\tau')}}{\Omega_k^2+\omega^2} =\frac{\lambda_k^2}{2 m_k \Omega_k}e^{-\Omega_k|\tau-\tau'|}.
\end{align}
Performing the sum over $k$ then yields
\begin{align}
    \mathcal{D}(\tau,\tau')&=\int d\Omega \,e^{-\Omega|\tau-\tau'|}\sum_k\frac{\lambda_k^2}{2 m_k \Omega_k} \delta(\Omega-\Omega_k),
\end{align}
where we recognize the spectral function from Eq.~\ref{eq:spectral_function} as $J(\Omega)=\frac{\pi}{2}\sum_k\frac{\lambda_k^2}{m_k \Omega_k} \delta(\Omega-\Omega_k)=\alpha \frac{\pi \Omega_D^{1-s}}{\Gamma(1+s)} \Omega^s \text{ for }\Omega\in [0,\Omega_D]$. This leads to
\begin{align}
    \mathcal{D}(\tau,\tau')&=\alpha \frac{\Omega_D^{1-s}}{\Gamma(1+s)}\int_0^{\Omega_D} d\Omega\,\Omega^s e^{-\Omega|\tau-\tau'|}\nonumber\\
    &=\frac{\alpha \Omega_D^{1-s}}{|\tau-\tau'|^{1+s}}\frac{\int_0^{\Omega_D|\tau-\tau'|} dx\,x^s e^{-x}}{\Gamma(1+s)}.
\end{align}
For $|\tau-\tau'|\gg \tau_c=1/\Omega_D$, one can approximate the integral as $\int_0^\infty dx x^s e^{-x}=\Gamma(1+s)$. Thus
\begin{equation}
    \mathcal{D}(\tau,\tau')=\frac{\alpha \tau_c^{s-1}}{|\tau-\tau'|^{1+s}},
\end{equation}
which is the expression given in the main text.

\section{Perturbative RG analysis}\label{appendix:RG_OPE}
The goal of this appendix is to compute the RG equations Eqs.~\ref{eq:RG_eq_1},\ref{eq:RG_eq_2},\ref{eq:RG_eq_3},\ref{eq:RG_eq_4} by means of the operator product expansion (OPE).

\subsection{Derivation of the useful OPEs}
The operator product expansion (OPE) is a series expansion of a product of two nearby fields in terms of local fields. This is done on normal ordered operators to avoid any divergence not coming from the two operators being pushed together. For our problem, we will need the following OPEs
\begin{align}
    :e^{2ip\phi(r)}&::e^{2ip\phi(r')}:=a^{2p^2K}:e^{4ip\phi(r)}:,\\
    :e^{2ip\phi(r)}&::e^{-2ip\phi(r')}:=\frac{:e^{2ip(\phi(r)-\phi(r'))}:}{|\delta r|^{2p^2K}}\nonumber\\
    &\hspace{-1cm}=\frac{:1+2ip\delta r\cdot \nabla \phi(r)-2p^2(\delta r\cdot \nabla \phi(r))^2:}{|\delta r|^{2p^2K}},
\end{align}
which imply
\begin{align}
\label{eq:OPE_cos}
    :\cos [2p\phi(r)]:&:\cos [2p\phi(r')]:=\frac{a^{2p^2K}}{2}:\cos[4p\phi(r)]:\nonumber\\
    &-p^2\frac{:(\delta r\cdot \nabla \phi(r))^2):}{|\delta r|^{2p^2K}}+...,\\
    :\cos[4\phi(r)]:&:\cos[2\phi(r')]:=\frac{:\cos[2\phi(r)]:}{2a^{4K}}+...\label{eq:OPE_cos_24}
\end{align}
where $r=(x,u\tau)$, $r'=(x',u\tau')$, $\delta r=r'-r$. These relations are easily proven using the identity $:e^{i2p\phi(r)}:=\frac{e^{i2p\phi(r)}}{\langle e^{i2p\phi(r)} \rangle}=\frac{e^{i2p\phi(r)}}{a^{p^2 K}}$, where $\langle \cdot \rangle$ is the average with respect to the Gaussian action $S_{\rm LL}$.

\subsection{Renormalization group using the OPE}
The interacting part of the action in Eq.~\ref{eq:bosonized_action} can be rewritten in terms of normal ordered fields as
\begin{align}
    S_{\rm int}=&S_g+S_\alpha\nonumber\\
    =&-\frac{g u}{2 \pi^2 a^{2-4K}}\int dx d\tau :\cos \left[ 4\phi(x,\tau)\right]:\nonumber\\
    &-\frac{\alpha u^{1-s}}{2 \pi^2 a^{2-s-2K}}\underset{|\tau-\tau'|>\tau_c}{\int dx d\tau d\tau'}\nonumber\\
    &\times \frac{:\cos \left[ 2 \phi(x,\tau)\right]::\cos \left[ 2 \phi(x,\tau')\right]:}{|\tau-\tau'|^{1+s}}.
\end{align}
One then expands the partition function up to order $\mathcal{O}(\alpha,g^2,\alpha g)$ such that
\begin{align}
    Z&=\int \mathcal{D}\phi e^{-S_{\rm LL}[\phi]-S_g[\phi]-S_\alpha[\phi]}\nonumber\\
    &=Z_{\rm LL}\Big[1- \langle S_g\rangle(a)-\langle S_\alpha\rangle(a)\nonumber\\
    &\hspace{1cm}+\frac{\langle S_g^2\rangle(a)}{2}+\langle S_\alpha S_g\rangle(a) \Big],
\end{align}
where the cutoff dependence has been made explicit. We now perform a rescaling of the lattice spacing $a\to a'=a(1+ dl)$ (so $\tau_c \to \tau_c'=\tau_c(1+dl)$) and ask how the couplings should vary to preserve the partition function $Z$. $Z_{\rm LL}$ being an RG fixed-point, we only need to consider the variations of the averaged interacting terms. Keeping only the first order in $dl$, $\langle S_g\rangle(a')$ becomes
\begin{align}
\langle S_g\rangle (a')&=\langle S_g\rangle (a)\left[ 1+\frac{dg}{g}+(4K-2)dl\right].
\end{align}
The next term $\langle S_\alpha\rangle (a')$ requires a bit more work. We start by splitting the imaginary-time integral as \hspace{0.5cm} $\underset{|\tau-\tau'|>\tau_c}{\int [...]} - \underset{\tau_c'>|\tau-\tau'|>\tau_c}{\int [...]} $. The first term is simply given by
\begin{align}
\langle S_\alpha\rangle (a)\left[ 1+\frac{d\alpha}{\alpha}+(2K+s-2)dl\right],
\end{align}
while the second term, which involves only fields at very close positions, can be evaluated using the OPE \ref{eq:OPE_cos} and gives
\begin{align}
    dl&\Big[\frac{\alpha(a)u}{2 \pi^2 a^{2-4K}}\int dx d\tau \langle:\cos\left[4\phi(x,\tau)\right]:\rangle\nonumber\\
    &-\frac{\alpha(a)}{\pi^2 u}\int dx d\tau\langle:(\partial_\tau \phi(x,\tau))^2:\rangle\Big].
\end{align}
The third term, namely $\frac{1}{2}\langle S_g^2\rangle(a')$, reads
\begin{align}
    \frac{\langle S_g^2\rangle(a')}{2}=\frac{g^2a'^{8K-4}}{8\pi^4}&\underset{|r-r'|>a'}{\int dr dr'}\\
    &\times\langle :\cos 4\phi(r)::\cos 4\phi(r'):\rangle.\nonumber
\end{align}
For this term, one again splits the limits of the integral into $\int \underset{|r-r'|>a}{dr dr'}[...]-\int \underset{a'>|r-r'|>a}{dr dr'}[...].$
The first part is easily tractable while the second requires the use of the OPE \ref{eq:OPE_cos}. This yields
\begin{align}
    \frac{\langle S_g^2\rangle(a')}{2}=&\frac{\langle S_g^2\rangle(a)}{2}\left[1+\frac{2dg}{g}+(8K-4)dl \right]\nonumber\\
    &+dl\frac{g^2}{2\pi^3}\int dr :(\nabla \phi(r))^2:.
\end{align}
The last term $\langle S_g S_\alpha \rangle(a')$ can be, yet again, separated into a simple part and a part that requires the OPE \ref{eq:OPE_cos_24} to give
\begin{align}
    \langle S_g S_\alpha\rangle(a')=&\langle S_g S_\alpha\rangle(a)\Big[1+\frac{dg}{g}+\frac{d\alpha}{\alpha}\\
    &+(6K+s-4)dl\Big]+dl \frac{g}{\pi}\langle S_\alpha \rangle(a).\nonumber
\end{align}
Putting everything together, one arrives at
\begin{align}
    &\langle S_g\rangle (a')+\langle S_\alpha\rangle (a')-\frac{\langle S_g^2\rangle}{2} (a')-\langle S_g S_\alpha\rangle(a')\nonumber\\
    =&\langle S_g\rangle (a)\left[1+\frac{dg}{g}+(4K-2)dl-\frac{\alpha}{g}dl\right]\nonumber\\
    &+\langle S_\alpha\rangle (a)\left[1+ \frac{d\alpha}{\alpha}+(2K+s-2)dl-\frac{g}{\pi}dl\right]\nonumber\\
    &-\frac{\langle S_g^2\rangle(a)}{2}\left[1+\frac{2dg}{g}+(8K-4)dl \right]\nonumber\\
    &-\langle S_g S_\alpha\rangle(a)\left[1+\frac{dg}{g}+\frac{d\alpha}{\alpha}+(6K+s-4)dl\right]\nonumber\\
    &-dl\frac{\alpha(a)}{\pi^2 u}\int dx d\tau\langle:(\partial_\tau \phi(x,\tau))^2:\rangle\nonumber\\
    &-dl\frac{g^2}{2\pi^3}\int dr :(\nabla \phi(r))^2:.
\end{align}
Upon imposing that the partition function $Z$ remains unchanged, the RG equations for $\alpha$ and $g$ can be directly read off. Those for $K$ and $u$ are found by re-exponentiating the remaining quadratic terms. In the end, one finds
\begin{align}
    \frac{d}{dl}&\frac{u}{K}=\frac{g^2u}{\pi^2},\\
    \frac{d}{dl}&\frac{1}{uK}=\frac{2\alpha}{\pi u}+\frac{g^2}{\pi^2u},\\
    \frac{d}{dl}&g=(2-4K)g+\alpha,\\
    \frac{d}{dl}&\alpha=(2-s-2K)\alpha+\frac{g\alpha}{\pi}.
\end{align}

\section{Gap closure in the ohmic case}
\label{appendix:p_ohmic}
This appendix derives the parameter $p$ of the gap closure $\Delta\sim \exp(-Ct^{-p})$ in the ohmic case ($s=1$), with $t$ the distance in coupling space to the separatrix plane. The following derivation is based on \cite{Nelson1978}.

\subsection{Analytic derivation}
To extract the behaviour of the microscopic gap $\Delta(0)$ near the transition, one notices that the renormalized gap is simply $\Delta(l)=e^l \Delta(0)$ because of its scaling dimension. If one starts the RG flow in the dissipative phase, the couplings $g(l)$ and $\alpha(l)$ will become of order $\mathcal{O}(1)$ after a renormalization time $l^\star$. This corresponds to a renormalized gap $\Delta(l^\star)$ of order $\mathcal{O}(1)$, so $\Delta(0)\sim e^{-l^\star}$. The goal of the following argument is thus to extract this time $l^\star$.

\begin{figure}
    \centering
    \includegraphics[width=7cm]{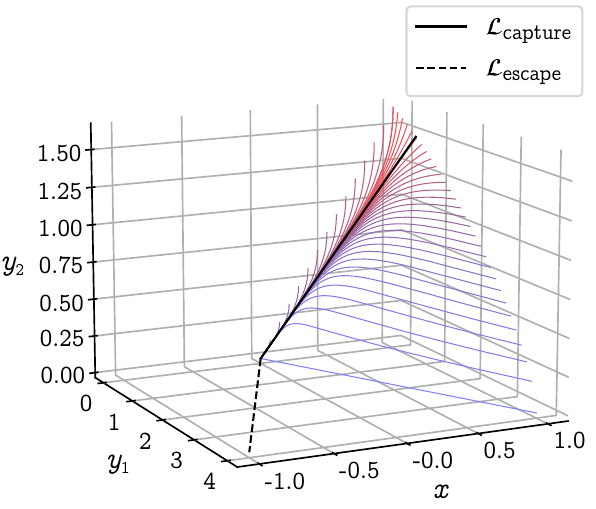}
    \caption{The blue to red lines are trajectories which sit on the separatrix manifold between both phases. The large $x$ side is the LL, while the small and negative $x$ side is the AFM. All trajectories starting near the separatrix are quickly attracted to the the line $\mathcal{L}_{\rm capture}$. Once they arrive near the origin, they tend to follow $\mathcal{L}_{\rm escape}$.}
    \label{fig:ohmic_RG}
\end{figure}

Let us start from the RG equations \ref{eq:RG_eq_1}-\ref{eq:RG_eq_4}. Since $K_c=1/2$ at $s=1$, one sets $K=1/2+x$ to study the vicinity of the critical point. We also introduce $y_1=g/\pi$ and $y_2=\sqrt{\alpha/\pi}$. To leading order in $x,y_1,y_2$, the RG equations then read
\begin{align}
    \frac{d}{dl}&x=-\frac{y_1^2+y_2^2}{4},\label{eq:RG_x}\\
    \frac{d}{dl}&y_1=-4x y_1+y_2^2,\label{eq:RG_y1}\\
    \frac{d}{dl}&y_2=-x y_2+\frac{y_1 y_2}{2},\label{eq:RG_y2}
\end{align}
where we have dropped the equation for $u$ since it does not feedback into the other equations. Let us consider a trajectory starting near the separatrix (which is here a 2D manifold, see Fig.~\ref{fig:ohmic_RG}) and in the dissipative phase. Such a trajectory starts by shooting towards the origin $(0,0,0)$, then drastically slows down near this point, and finally escapes towards $(-\infty,\infty,\infty)$. This corresponds to a capture time $l^\star_1$, followed by a transition time $l^\star_2$, and an escape time $l^\star_3$, which add up to a total dissipative time $l^\star=l^\star_1+l^\star_2+l^\star_3$.

From numerical simulations (see Fig.~\ref{fig:ohmic_RG}), it appears that all trajectories near the separatrix converge rapidly to a common line $\mathcal{L}_{\rm capture}$. Plugging the ansatz $y_1=m_1x$ and $y_2=m_2x$ into Eqs.~\ref{eq:RG_x}-\ref{eq:RG_y2} shows that
\begin{equation}
    \mathcal{L}_{\rm capture}: y_1=\frac{2}{3}x, y_2=\frac{\sqrt{20}}{3}x.
\end{equation}
Since $y_1$ and $y_2$ remain finite throughout this process, the time it takes to reach $\mathcal{L}_{\rm capture}$ will be of order $\mathcal{O}(t^0)\ll l^\star$ and the distance $t$ to the separatrix will only be rescaled by a finite prefactor. We can therefore assume that our trajectory starts at a point close to $\mathcal{L}_{\rm capture}$ with a distance $t$ to the separatrix. Once the trajectory has come close to the origin, it escapes along the line
\begin{equation}
    \label{eq:L_escape}
    \mathcal{L}_{\rm escape}: y_1=-4x, y_2=0.
\end{equation}

\paragraph{Capture time} During the capture phase the trajectory stays close to the capture line $\mathcal{L}_{\rm capture}$, hence the notation $y_1=\frac{2}{3}x+D_1$, $y_2=\frac{\sqrt{20}}{3}x+D_2$ with $D_1,D_2$ the deviations from $\mathcal{L}_{\rm capture}$. At leading order in the deviations, Eq.~\ref{eq:RG_x} becomes
\begin{align}
    \label{eq:x_sol_capture}
    \frac{d}{dl}&x=-\frac{2}{3}x^2  \Rightarrow x(l)=\frac{x_0}{1+\frac{2 x_0 l}{3}},
\end{align}
where we have set $x(l=0)=x_0$. Equations \ref{eq:RG_y1},\ref{eq:RG_y2} are then given to leading order by
\begin{equation}
    \frac{d}{dl}\begin{pmatrix} D_1 \\ D_2 \end{pmatrix}=\frac{x}{9}\begin{pmatrix} -34 & 7\sqrt{20}\\ 5\sqrt{20}/2 & 4 \end{pmatrix}\begin{pmatrix} D_1 \\ D_2 \end{pmatrix}.
\end{equation}
Diagonalizing the $2\times2$ matrix shows that
\begin{equation}
    \frac{d}{dl}\begin{pmatrix} D_+ \\ D_- \end{pmatrix}=x\begin{pmatrix} \lambda_+ & 0\\ 0 & \lambda_- \end{pmatrix}\begin{pmatrix} D_+ \\ D_- \end{pmatrix},
\end{equation}
with $\lambda_\pm=-\frac{5}{3}\pm\frac{\sqrt{79}}{3}$ and $D_\pm=\pm\frac56 \sqrt{\frac{5}{79}}D_1+\left(\frac12 \pm \frac{19}{6\sqrt{79}}\right)D_2$. Using Eq.~\ref{eq:x_sol_capture}, the deviations $D_\pm$ are found to be
\begin{align}
    \label{eq:D_sol_capture}
    &D_\pm(l)=D_\pm(l=0)\left( 1+\frac{x_0 l}{2}\right)^{3\lambda_\pm/2}.
\end{align}
Intuitively, $D_-$ is the deviation from $\mathcal{L}_{\rm capture}$ within the separatrix manifold while $D_+$ is the deviation outside of the manifold. Since, $\lambda_-<0$ and $\lambda_+>0$, $D_-$ quickly becomes negligible compared to $D_+$ and is therefore just dropped. Moreover, $D_+$ being the deviation from the separatrix, its initial condition is $D_+(l=0) \sim t$.

The capture phase stops at a time $l^\star_1$ such that $D_+(l^\star_1)\sim x(l^\star_1)$. Using the expressions for $D_+(l)$ and $x(l)$ derived in Eqs.~\ref{eq:x_sol_capture},\ref{eq:D_sol_capture},
\begin{equation}
    l^\star_1 \sim t^{-\frac{1}{3\lambda_+/2 + 1}}.
\end{equation}
From the previous equations, it also follows that $x(l^\star_1) \sim D_+(l^\star_1) \sim 1/l^\star_1$.

\paragraph{Transition time} Starting at $l^\star_1$, the trajectory is stuck for some time $l^\star_2$ around the origin. During this time, $y_1$ and $y_2$ remain rather constant and $x$ goes from $+x(l^\star_1)$ to $x(l^\star_1+l^\star_2)\sim-x(l^\star_1)$. Since $\frac{dx}{dl}= -(y_1^2+y_2^2)/4 \sim -1/{l^\star_1}^2$ and $x(l^\star_1)\sim 1/l^\star_1$, this entire process takes a time $l^\star_2=2 x(l^\star_1)/(dx/dl)\sim l^\star_1$.

\paragraph{Escape time} For $l>l^\star_{12}=l^\star_1+l^\star_2$, the trajectory follows $\mathcal{L}_{\rm escape}$. Looking at Eq.~\ref{eq:L_escape}, the adapted deviations are defined through $y_1=-4x+D_1$, $y_2=0+D_2$. To leading order, Eq.~\ref{eq:RG_x} reduces to (recall that $x(l^\star_{12})<0$)
\begin{align}
    \frac{d}{dl}&x=-4x^2\Rightarrow x(l)=\frac{x(l^\star_{12})}{1+4(l-l^\star_{12})x(l^\star_{12})},
\end{align}
The escape time is reached when $x(l)$ becomes of order $\mathcal{O}(1)$, that is when $l^\star_3=l^\star-l^\star_{12}\sim 1/x(l^\star_{12})$, i.e. $l^\star_3 \sim l^\star_1$.

Adding up $l^\star_1,l^\star_2,l^\star_3$ shows that $l^\star \sim t^{-p}$ with $p=\frac{1}{3\lambda_+/2 + 1}=\frac{\sqrt{79}+3}{35}\simeq 0.33966269\dots$.

\begin{figure}[h!]
    \centering
    \includegraphics[width=7cm]{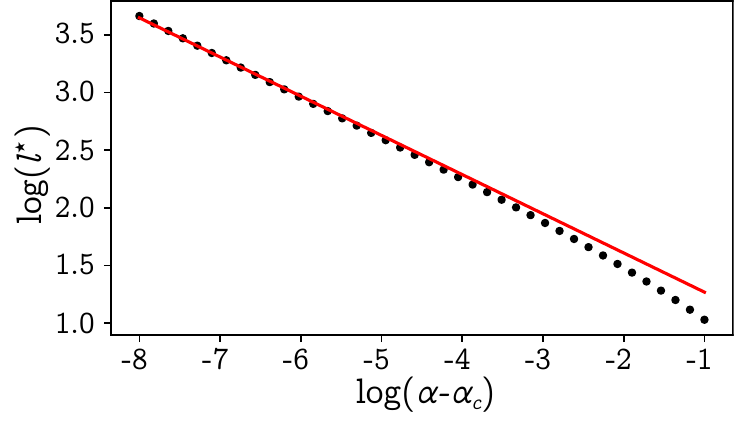}
    \caption{Scaling of the total dissipative time $l^\star$ for trajectories starting close to the transition point $\alpha_c$ at $K=1$ and $g=0.5$. The dots are the numerical data points while the line is a fit using the ansatz $\log(l^\star)=-0.3396\times\log(\alpha-\alpha_c)+C$.}
    \label{fig:fit_exponent}
\end{figure}

\subsection{Numerical confirmation}
The RG equations for $K,g,\alpha$ can be numerically integrated for any initial condition. For the initial values $K=1$ and $g=0.5$, we vary $\alpha$ to get close to the transition point $\alpha_c$ and measure the associated total dissipative time $l^\star$. This time is defined as the time needed for the simulation to reach the plane $K=0.1$. The results are shown in Fig.~\ref{fig:fit_exponent}. It appears that the data fits quite nicely with our ansatz $\log(l^\star)=-0.3396\times\log(\alpha-\alpha_c)+C$ for $\alpha-\alpha_c \ll 1$.

\section{Gap equation from the variational method}
\label{appendix:gap_eq}
We wish to find a solution to the self-consistent variational equation Eq.~\ref{eq:selfconsist} using the following ansatz $G^{-1}_{\rm AFM}(q,\omega_n)=\frac{1}{\pi K}\left[uq^2+ \frac{\omega_n^2}{u}+\frac{\Delta^2}{u}\right]$. Setting $\omega_n=q=0$ in Eq.~\ref{eq:selfconsist} yields
\begin{align}
    \label{eq:delta_eq_appendix}
    \frac{\Delta^2}{\pi K u}&=\frac{8g}{\pi^2a^2}\exp \left[-2K\int  \frac{d\omega_n'}{\sqrt{{\omega_n'}^2+\Delta^2}}\right]\\
    &\hspace{-1cm}+\frac{4\alpha u^{-s}}{\pi^2a^{2-s}}\int_{\tau_c}^\infty \frac{d\tau}{\tau^{1+s}}\exp \left[-K\int d\omega_n' \frac{1+\cos \omega_n' \tau}{\sqrt{{\omega_n'}^2+\Delta^2}} \right],\nonumber
\end{align}
where we have done both integrals over $q$ in the exponentials. The remaining integrals over $\omega_n'$ are UV divergent and are regulated by reintroducing the imaginary-time cutoff $\tau_c=a/u$ (thus $1/\tau_c$ in Matsubara frequency space). In the limit of $\Delta \tau \ll 1$, one can then find that $\int_0^{1/\tau_c}  \frac{d\omega_n'}{\sqrt{{\omega_n'}^2+\Delta^2}}=-\log(\Delta \tau_c)$, and $\int_0^{1/\tau_c} d\omega_n' \frac{1+\cos \omega_n' \tau}{\sqrt{{\omega_n'}^2+\Delta^2}} =-\log(\Delta^2\tau_c\tau)-\gamma_E$,
with $\gamma_E$ Euler's gamma constant. Plugging this back into Eq.~\ref{eq:delta_eq_appendix} gives
\begin{align}
    \frac{\Delta^2}{\pi K}=&\frac{8g}{\pi^2\tau_c^{2-4K}} \Delta^{4K}\\
    &+\frac{4\alpha}{\pi^2\tau_c^{2-s-2K}}\int_{\tau_c}^\frac{1}{\Delta} \frac{d\tau}{\tau^{1+s-2K}}\Delta^{4K}e^{2K\gamma_E},\nonumber
\end{align}
which, after doing the integral over $\tau$ and rearranging the terms, leads to

\begin{align}
    \frac{(\Delta\tau_c)^{2-4K}}{K}=&g\frac{8}{\pi}+\alpha\frac{4 e^{2K\gamma_E}}{\pi}\frac{(\Delta\tau_c)^{s-2K}-1}{2K-s}.
\end{align}
This is Eq.~\ref{eq:delta_phase_trans} in the text.

\section{Variational method vs RG}
\label{appendix:var_vs_RG}
\begin{figure}
    \centering
    \includegraphics[width=7.5cm]{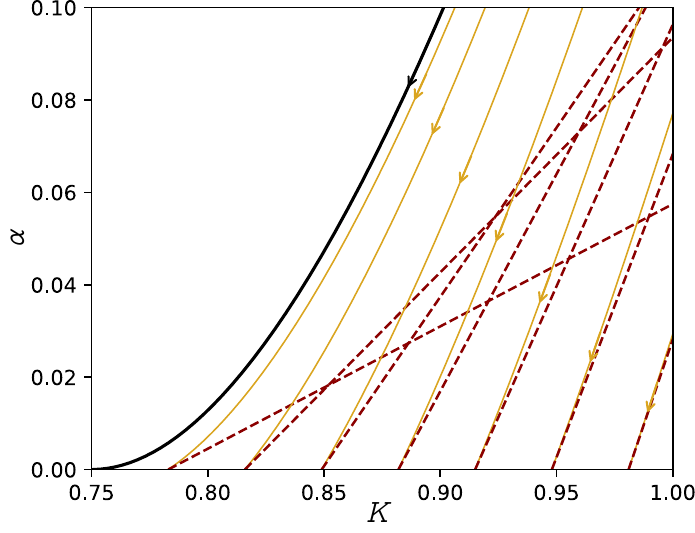}
    \caption{RG flow in the LL phase of the couplings $\alpha$ and $K$ according to the actual RG equations (solid gold lines) and the linearized equations (dashed red lines). This plot is for $s=0.5$.}
    \label{fig:alphaK_RGlinflow}
\end{figure}

This section shows how the ansatz $G^{-1}_{\rm var}(q,\omega_n)=\frac{1}{\pi K_r}\left[u_rq^2+ \frac{\omega_n^2}{u_r}+\frac{\Delta^2}{u_r}\right]$ fails at predicting the correct renormalized coefficients close to the transition. For simplicity and to compare the result with the perturbative RG, we focus on the LL phase where $\Delta_r=0$. Taking the second derivative with respect to $q$ of Eq.~\ref{eq:selfconsist} shows that $\frac{u_r}{K_r}=\frac{u}{K}$ so we can express both Luttinger parameters in terms of $\eta$ such that $K=K_r \sqrt{1+\eta}$, $u=u_r\sqrt{1+\eta}$. The ansatz thus becomes $G^{-1}_{\rm var}(q,\omega_n)=\frac{1}{\pi K}\left[uq^2+ \frac{\omega_n^2}{u}(1+\eta)\right]$ where $\eta$ is the only parameter to solve for. Taking the second derivative of Eq.~\ref{eq:selfconsist} with respect to $\omega_n$ yields
\begin{align}
    \eta=&\frac{\alpha K}{\pi \tau_c^{2-s}}\int_{\tau_c}^\infty \frac{d\tau}{\tau^{-1+s}}\\
    &\times \exp \left[-\frac{K}{\pi}\int dq' d\omega_n' \frac{1- \cos \omega_n' \tau }{u{q'}^2+\frac{{\omega_n'}^2}{u}(1+\eta) } \right].\nonumber
\end{align}
We now perform the integral over $q$ and regulate the diverging integral over $\omega_n'$ by adding the imaginary-time UV cutoff $\tau_c$ to obtain
\begin{align}
    \eta=&\frac{\alpha K}{\pi \tau_c^{2-s}}\int_{\tau_c}^\infty \frac{d\tau}{\tau^{-1+s}}\\
    &\times \exp \left[-2K_r\int_0^{1/\tau_c} d\omega_n' \frac{1- \cos \omega_n' \tau }{\omega_n'} \right].\nonumber
\end{align}
In the limit of large $\tau/\tau_c$, one proves the following identity
\begin{align}
\label{eq:artefact_RG_var}
    \int_0^{1/\tau_c}d\omega_n' \frac{1- \cos \omega_n' \tau }{\omega_n'}=\log(\tau/\tau_c)+\gamma_E+\mathcal{O}(\tau_c/\tau),
\end{align}
which implies that
\begin{align}\label{Eq_eta_ren}
    \eta=&\frac{\alpha K e^{-2K_r\gamma_E}}{\pi \tau_c^{2-s-2K_r}}\int_{\tau_c}^\infty \frac{d\tau}{\tau^{-1+s+2K_r}}.
\end{align}
Computing this integral and using $K=K_r\sqrt{1+\eta}$ leads to
\begin{align}
\label{eq:eta}
    \frac{\eta}{\sqrt{1+\eta}}=&\frac{\alpha K_r e^{-2K_r\gamma_E}}{\pi(2K_r+s-2)}.
\end{align}
Now comes the problem. From the argument made in section \ref{sec:var_method}, it is clear that the transition is given in terms of $K_r$ by $(K_r)_c=\max(1-s/2,1/2)$. This means that for $s<1$, Eq.~\ref{eq:eta} predicts a renormalization of $K$ and $u$ that diverges towards the transition as
\begin{align}
\label{eq:diverging_eta}
    \sqrt{\eta}\simeq &\frac{\alpha K_r e^{-2K_r\gamma_E}}{\pi(2K_r+s-2)}.
\end{align}
which is of course highly unphysical. This result can however be recovered from the RG using a very crude approximation: linearizing the RG flow about $\alpha=0$ (see fig.~\ref{fig:alphaK_RGlinflow}). To show how this recovers Eq.~\ref{eq:diverging_eta}, let us start by recalling the important RG equations for $s<1$ (see subsection \ref{subsec:RG_subohmic})
\begin{align}
    \frac{d}{dl}&K=-\frac{\alpha K^2}{\pi},\\
    \frac{d}{dl}&\alpha=(2-s-2K)\alpha.\label{eq:Kr_def}
\end{align}
In the LL phase, $\alpha$ flows to 0 and $K$ to $K_r$. This means that $K_r$ is given by $K_r=K+\int_K^{K_r} dK'=K+\int_\alpha^0 \frac{dK}{d\alpha}(\alpha') d\alpha'$. For the linearized RG flow shown in fig.~\ref{fig:alphaK_RGlinflow}, $\frac{dK}{d\alpha}(\alpha')=\frac{dK}{d\alpha}(\alpha' = 0)=\frac{K_r^2}{\pi(2K_r+s-2)}$. Plugging this back into Eq.~\ref{eq:Kr_def} and using $K=K_r\sqrt{1+\eta}$ yields
\begin{align}
    \sqrt{1+\eta}-1=&\frac{\alpha K_r}{\pi(2K_r+s-2)}.
\end{align}
\begin{figure*}[t!]
    \centering
    \includegraphics[width=1\linewidth, clip, trim=80 135 100 125]{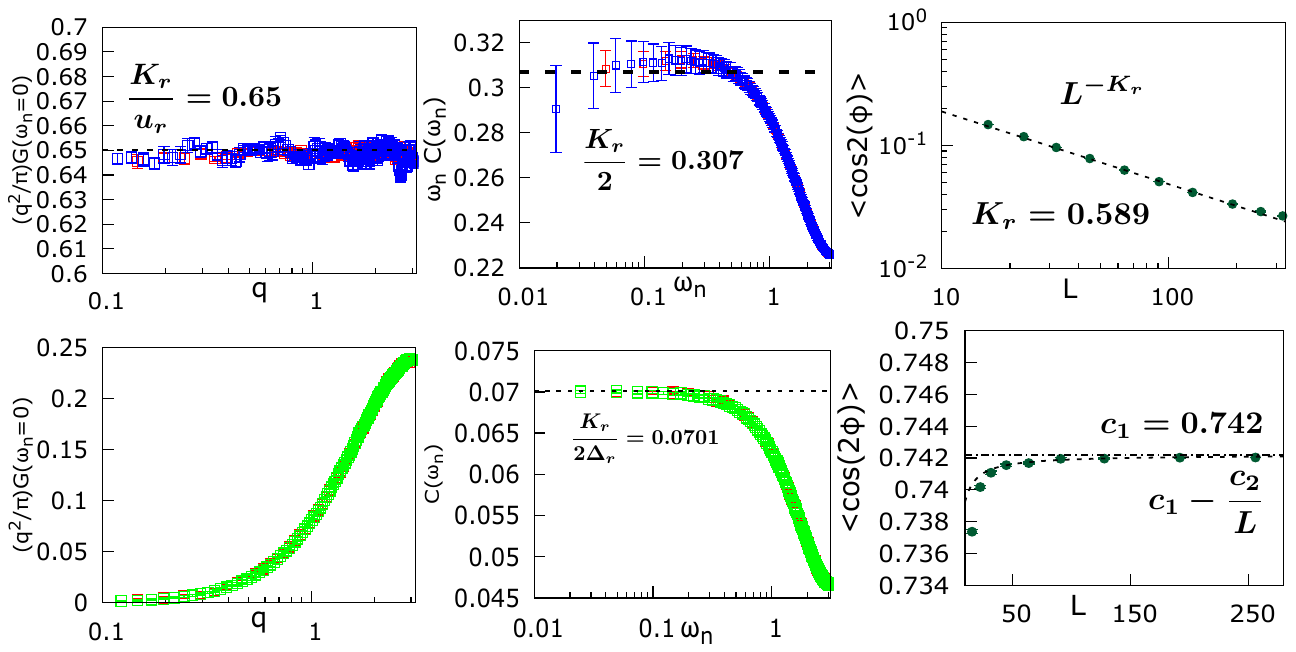}
    \caption{Numerical evaluation of observables for the commensurate dissipative spin chain. The values of the parameters are taken to be $s=1$ ($K_c=0.5$), $K=0.65$, $g=0.5$, $u=1$, and $\beta$ is scaled as $L$. For the \textit{left} and \textit{middle} column, red, green, and blue colors denote $L=128$, $L=256$ and $L=320$ respectively. All the quantities have been averaged over 10000-20000 configurations and the value of time-step $dt$ is $0.5$. (top) and (bottom) rows correspond to $\alpha=0.5$ (LL) and $\alpha=6$ (AFM), respectively. (left) Susceptibility $\chi$ as a function $q$. For $\alpha=0.5$, it stays finite and constant whereas for $\alpha=6$, it vanishes for small $q$. (middle) Behavior of $C(\omega_n)$ as a function of $\omega_n$. For $\alpha=0.5$, $\omega_n C(\omega_n)$ saturates to a constant $K_r/2=0.307$; for $\alpha=6$, $C(\omega_n)$ becomes a constant $K_r/(2\Delta_r)=0.07$. (right) Behavior of order parameter $\langle \cos \left[2\left(\phi-\phi_{\rm CoM} \right)\right] \rangle$  as a function of system size $L$. For $\alpha=0.5$, the order parameter decays as $L^{-K_r}$ with $K_r = 0.59$, whereas for $\alpha=6$, it increases and saturates to a constant as $c_1 - c_2/L$ with $c_1 = 0.742$ and $c_2=0.0296$.}
    \label{fig:ohmic_result}
\end{figure*}
Near the transition ($K_r\to(1-s/2)^+$), this equation becomes exactly Eq.~\ref{eq:diverging_eta} up to the numerical factor $e^{-2K_r\gamma_E}$ which is probably an artefact of the approximation made in Eq.~\ref{eq:artefact_RG_var}. The main takeaway message is that the variational method captures the RG flow linearized about $\alpha=0$, and thus breaks down close to the transition.
\begin{figure*}[t!]
    \centering
    \includegraphics[width=1\linewidth, clip, trim=60 135 150 125]{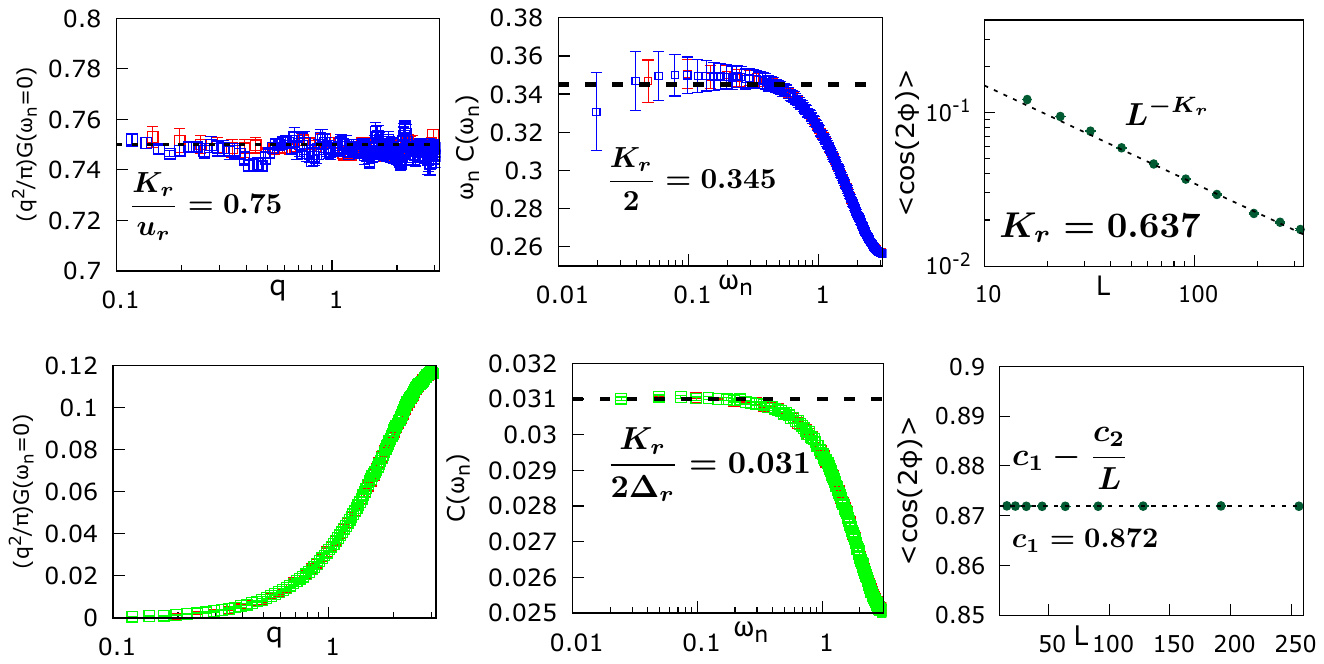}
    \caption{Numerical evaluation of observables for the commensurate dissipative spin chain. The values of the parameters are taken to be $s=1.5$ ($K_c=0.5$), $K=0.75$, $g=1$, $u=1$, and $\beta$ is scaled as $L$. For the \textit{left} and \textit{middle} column, red, green, and blue colors denote $L=128$, $L=256$ and $L=320$ respectively, and $\beta$ is scaled as $L$. All the quantities have been averaged over 10000-20000 configurations and the value of time-step $dt$ is $0.5$. (top) and (bottom) rows correspond to $\alpha=0.65$ (LL) and $\alpha=6$ (AFM), respectively. (left) Susceptibility $\chi$ as a function $q$. For $\alpha=0.65$, it stays finite and constant whereas for $\alpha=6$, it vanishes for small $q$. (middle) Behavior $C(\omega_n)$ as a function of $\omega_n$. For $\alpha=0.65$, $\omega_n C(\omega_n)$ saturates to a constant $K_r/2=0.345$; for $\alpha=6$, $C(\omega_n)$ becomes a constant $K_r/(2\Delta_r)=0.031$. (right) Behavior of order parameter $\langle \cos \left[2\left(\phi-\phi_{\rm CoM} \right)\right] \rangle$  as a function of system size $L$. For $\alpha=0.65$, the order parameter decays as $L^{-K_r}$ with $K_r = 0.637$, whereas for $\alpha=5$, it increases and saturates to a constant as $c_1 - (c_2/L)$ with $c_1 = 0.872$ and $c_2=0$.}
    \label{fig:superohmic_result}
\end{figure*}

\section{A simple argument}
\label{appendix:var_non_selfcons}

In this section we show that the previous arguments allow to correctly capture the phase transition if the renormalisation of $K$ is taken into account in a simpler manner, namely without resorting to a self consistent computation. This approximation should correspond to assuming that all the way down to $K_r=K_c$ the computation of $K_r$ is perturbative in $\alpha$. The computation amounts to replacing, in Eq.~\ref{Eq_eta_ren}, the value of $K_r$ with $K$ as
\begin{equation}
\eta = \frac{\alpha K C}{2\pi(K-K_c)},
\end{equation}
with $C=e^{-2K \gamma_E}$, and then using $\eta$'s definition
\begin{equation}
\eta=\left(\frac{K}{K_r}\right)^2 - 1,
\end{equation}
which, for $K_r=K_c$, recovers
\begin{equation}
\left(K-K_c\right)^2 \propto \alpha,
\end{equation}
which is the parabolic shape of the transition predicted by RG.

\section{Derivation of the order parameter}\label{App:OP}
In this appendix, we show the calculation of the order parameter $\langle \cos \left( 2\left(\phi - \phi_{\rm CoM} \right) \right) \rangle$ using our Gaussian variational ansatz. The behavior of this quantity in the LL has already been calculated  in \cite{Sap_subohmic} (appendix A shows that $\langle \cos \left( 2\left(\phi - \phi_{\rm CoM} \right) \right) \rangle\propto L^{-K_r}$) so we concentrate on the AFM phase defined by the propagator $G_{\rm AFM}^{-1}(q,\omega_n)=\frac{1}{\pi K}(uq^2+\frac{\omega_n^2}{u}+\frac{\Delta_r^2}{u_r})$. Since we work within our Gaussian variational theory, one writes $\langle \cos 2 \left(\phi - \phi_{\rm CoM} \right) \rangle = \exp ( -2\langle (\phi-\phi_{\rm CoM})^2 \rangle )$ where $\langle (\phi-\phi_{\rm CoM})^2\rangle$ can be broken up as
\begin{align}
&\langle (\phi-\phi_{\rm CoM})^2 \rangle = \frac{1}{\beta L} \left[ \sum\limits_{q \neq 0} G_{\rm AFM}(q,0)\right.\nonumber\\
+&  \sum\limits_{\omega_n \neq 0} G_{\rm AFM}(0,\omega_n) + \left. \sum\limits_{\omega_n \neq 0, q \neq 0} G_{\rm AFM}(q,\omega_n) \right].
\end{align}
The first two terms can be computed for finite $\beta$ and $L$ using Matsubara sum techniques. This leads to
\begin{align}
\sum_{q \neq 0} G_{\text{AFM}}(q,0) =& \frac{\pi K_r L}{2 \Delta_r} \coth \left( \frac{\Delta_r L}{2 u_r} \right) - \frac{\pi K_r u_r}{\Delta_r^2},
\end{align}
and
\begin{align}
\sum_{\omega_n \neq 0} G_{\text{AFM}}(0,\omega_n) =& \frac{\pi K_r u_r \beta}{2\Delta_r} \coth \left( \frac{\Delta_r \beta}{2} \right) - \frac{\pi K_r u_r}{ \Delta_r^2}.
\end{align}
The third term is then replaced by its continuous limit ($L\to \infty,\beta\to \infty$) as the finite size effects are taken care of in the first two sums. Thus
\begin{align}
\frac{1}{\beta L}  &\sum_{\omega_n \neq 0, q \neq 0} G_{\rm AFM}(q,\omega_n) \nonumber\\
=& \frac{K_r}{4\pi} \underset{\sqrt{(u_r q)^2+\omega_n^2}<\frac{1}{\tau_c}}{\int d\omega_n dq} \frac{1}{u_r q^2 + \frac{\omega_n^2}{u_r} + \frac{\Delta_r^2}{u_r}} \nonumber\\
=& -\frac{K_r}{4} \log \left(\frac{(\tau_c \Delta)^2}{1+(\tau_c \Delta)^2}\right),
\label{eq:gap_OP}
\end{align}
where we have introduced the UV cutoff $\tau_c$. In our simulations, we take $L=\beta$ and then $L \to \infty$. Taking this same scaling, a large $L$ expansion of the order parameter reads
\begin{align}
\langle \cos (2 (\phi-\phi_{\rm CoM})) \rangle \simeq c_1 - \frac{c_2}{L} + \frac{c_3}{L^2},
\end{align} 
where $c_1 = \left( \frac{\Delta \tau_c}{\sqrt{1+(\Delta \tau_c)^2}}\right)^{K_r}$, $c_2 = c_1\frac{\pi K_r}{\Delta_r} \left( 1+ u_r \right)$, and $c_3 = c_1\frac{\pi K_r}{\Delta_r^2} \left[ 4 u_r + \pi K_r \left(1+u_r \right)^2 \right]$ respectively. We observe that the order parameter monotonically increases to a constant with a leading finite size dependence of $1/L$. This is a signature of the gapped phase, as a fractional phase would have finite size scaling $\propto L^{\frac{s}{2}-1}$, and a Luttinger liquid would have a $\propto L^{-K_r}$ behavior.

\section{Additional numerical results}\label{App:more_num_res}
In this appendix, we provide additional numerical results that fortify the analytical claims made in the main article. As discussed in section \ref{sec:num_res}, we calculate the same correlation functions ($\chi$, $C(\omega_n)$, and $\langle \cos \left[2\left(\phi-\phi_{\text{CoM}} \right) \right]\rangle$) for ohmic and superohmic baths. The results are provided in fig.~\ref{fig:ohmic_result} and fig.~\ref{fig:superohmic_result}, where numerical results are given for ohmic ($s=1$) and superohmic ($s=1.5$) baths respectively, in a similar fashion as in section \ref{sec:num_res} of the main article. These plots confirm that the phase transition occurs for any value of $s \in (0,2)$ and that the ordered phase is always an AFM phase. The values of the relevant simulation parameters and numerically extracted parameters are mentioned in the figure captions. 

\bibliography{refs}

\end{document}